%% file: V339Del_MNRAS.tex
\documentclass[a4paper,fleqn,usenatbib]{mnras}

\usepackage[T1]{fontenc}
\usepackage{ae,aecompl}

\usepackage{graphicx}	
\usepackage{amsmath}	
\usepackage{amssymb}	

\usepackage{pdflscape}
\usepackage{lscape,graphicx,pifont}
\usepackage{times}
\usepackage{natbib}

\usepackage{xtab}
\usepackage{longtable}

\newcommand{\Msun}{\mbox{\,M$_\odot$}}
\newcommand{\Lsun}{\mbox{\,L$_\odot$}}
\newcommand{\vunit}{\mbox{\,km\,s$^{-1}$}}
\newcommand{\mic}{\mbox{$\,\mu$m}}
\newcommand{\pion}[2]{{#1}\,{\sc {#2}}}
\newcommand{\fion}[2]{[{#1}\,{\sc {#2}}]}
\newcommand{\Ne}{\mbox{$n_{\rm e}$}}
\newcommand{\Ni}{\mbox{$n_{\rm i}$}}
\newcommand{\Te}{\mbox{$T_{\rm e}$}}
\newcommand{\ltsimeq}{\raisebox{-0.6ex}{$\,\stackrel
        {\raisebox{-.2ex}{$\textstyle <$}}{\sim}\,$}}
\newcommand{\gtsimeq}{\raisebox{-0.6ex}{$\,\stackrel
        {\raisebox{-.2ex}{$\textstyle >$}}{\sim}\,$}}

\newcommand{\vd}{\mbox{V339~Del}}

\newcommand{\chemone}{\raisebox{0.03cm}{$-$}} 

\title[{Dust shell of V339 Del}]{Rise and fall of the dust shell of
the classical nova V339 Delphini}

\author[A. Evans et al.]
{A. Evans$^1$\thanks{E-mail: a.evans@keele.ac.uk}, D. P. K. Banerjee$^2$, 
R. D. Gehrz$^{3}$, V. Joshi$^{2}$,
N. M. Ashok$^{2}$, \newauthor
V. A. R. M. Ribeiro$^{4}$,
M. J. Darnley$^{5}$, 
C. E. Woodward$^3$,
D. Sand$^{6}$, 
G. H. Marion$^{7}$, 
\newauthor
T. R. Diamond$^{8}$,   
S. P. S. Eyres$^{9}$, 
R. M. Wagner$^{10}$,
L. A. Helton$^{11}$, 
S. Starrfield$^{12}$, \newauthor
D. P. Shenoy$^3$,
J. Krautter$^{13}$, 
W. D. Vacca$^{11}$, %
M. T. Rushton$^{14}$ \\
\mbox{}\\
$^{1}$Astrophysics Group, Keele University, Keele, Staffordshire, ST5 5BG, UK\\ 
$^{2}$Physical Research Laboratory, Ahmedabad 380009, India\\  
$^{3}$Minnesota Institute for Astrophysics, School of Physics \& Astronomy
116 Church Street SE, University of Minnesota, Minneapolis, MN 55455, USA\\ 
$^{4}$Department of Astrophysics/IMAPP, Radboud University, P.O. Box 9010, 6500 GL Nijmegen, The Netherlands\\
and Department of Physics and Astronomy, Botswana International University of Science and Technology, Private Bag 16, Palapye, Botswana \\ 
$^{5}$Astrophysics Research Institute, Liverpool John Moores University, IC2 Liverpool Science Park, Liverpool, L3 5RF, UK\\ 
$^{6}$Department of Physics, Texas Tech University, Lubbock, TX 79409, USA\\ 
$^{7}$University of Texas at Austin, 1 University Station C1400, Austin, TX 78712-0259, USA\\ 
$^{8}$NASA Goddard Space Flight Center, Greenbelt, MD 20771, USA \\ 
$^{9}$Jeremiah Horrocks Institute, University of Central Lancashire, Preston PR1 2HE, UK\\ 
$^{10}$Department of Astronomy, The Ohio State University, 140 West 18th Avenue, Columbus, OH 43210, USA \\
   and  LBT Observatory, University of Arizona, Tucson, AZ 85721-0065, USA \\ 
$^{11}$USRA-SOFIA Science Center, NASA Ames Research Center, Moffett Field, CA 94035, USA\\ 
$^{12}$School of Earth and Space Exploration, Arizona State University, Box 871404, Tempe, AZ 85287-1404, USA\\ 
$^{13}$Landessternwarte-Zentrum f\"ur Astronomie der Universit\"at, K\"onigstuhl, D-69117 Heidelberg, Germany\\ 
$^{14}$Astronomical Institute of the Romanian Academy, Str. Cutitul de Argint 5, Bucharest, 040557, Romania \\ 
}

\date{Version SUB (\today)}

\pubyear{2016}

\begin{document}
\label{firstpage}
\pagerange{\pageref{firstpage}--\pageref{lastpage}}
\maketitle

\begin{abstract}
We present infrared spectroscopy of the classical nova \vd, obtained over a $\sim2$~year period.
The infrared emission lines were initially symmetrical, with HWHM velocities of 525\vunit.
In later ($t\gtsimeq77$days, where $t$ is the time from outburst) spectra however, 
the lines displayed a distinct asymmetry, with a much stronger blue wing, possibly due 
to obscuration of the receding component by dust.
Dust formation commenced at $\sim$~day~34.75 at a condensation temperature of $1480\pm20$~K,
consistent with graphitic carbon. Thereafter the dust temperature declined with time as 
$T_{\rm d}\propto{t}^{-0.346}$, also consistent with graphitic carbon. The mass of dust initally
rose, as a result of an increase in grain size and/or number, peaked at $\sim$~day 100,
and then declined precipitously. This decline was most likely caused by
grain shattering due to electrostatic stress after the dust was exposed to X-radiation.
An Appendix summarises Planck Means for carbon, and
the determination of grain mass and radius for a carbon dust shell.
\end{abstract}




\section{Introduction}
\label{intro}
Classical nova (CN) eruptions are produced by a thermonuclear runaway
(TNR) on the surface of a white dwarf (WD) that has been accreting material from
a companion star in a semi-detached binary system
\citep*[see][for recent comprehensive reviews]{CN2,BASI,CT,SIH}.

Following the explosion, some $10^{-5}-10^{-4}$\Msun\ of material, enriched in
metals, is expelled at speeds of $\sim$~several hundred to $\sim$~several
thousand \vunit. CN explosions may occur on carbon-oxygen (CO) or oxygen-neon (ONe) WDs.
The latter give rise to ``fast'' novae \citep[see][for a definition of CN speed class]{PG,warner}
that are characterised by coronal emission, the prodution of little or no dust, and are 
over-abundant (relative to solar abundances) in C, N, O, Ne, Mg, and Al; they are likely major producers of $^{22}$Ne and 
$^{26}$Al \citep{helton}. On the other hand
CNe originating on CO WDs tend to be ``slow'' or ``moderate-speed'' CNe, and often produce copious
amounts of dust that is mainly carbonaceous.

Consequently, by ejecting gas and dust into the interstellar medium, CNe partake in the
chemical enrichment of the Galaxy and indeed, there is evidence that nova debris was present
when the Solar Nebula formed \citep{pepin,haenecour}.

Here we present ground-based infrared (IR) observations of the CN \vd\ (Nova Delphini 2013);
observations of this CN from the {\it Stratospheric Observatory For Infrared Astronomy} 
\citep[SOFIA;][]{young}, 
together with near-contemporaneous photometry and spectroscopy obtained at the Mt Abu Infrared
Observatory, India,
and photometry from the O'Brien Observatory in Marine on St~Croix, Minnesota, USA,
have been described by \cite{gehrz-v339}.
IR observations of CNe are reviewed by \cite{gehrz-CN}, \cite{BA-BASI}, \cite{EG-BASI} and \cite*{gehrz-CT}.
We also present an optical spectrum obtained at the Multiple Mirror Telescope (MMT), located on Mt Hopkins,
Arizona.

\begin{figure}
\begin{center}
\leavevmode
\includegraphics[angle=0,keepaspectratio,width=\columnwidth]{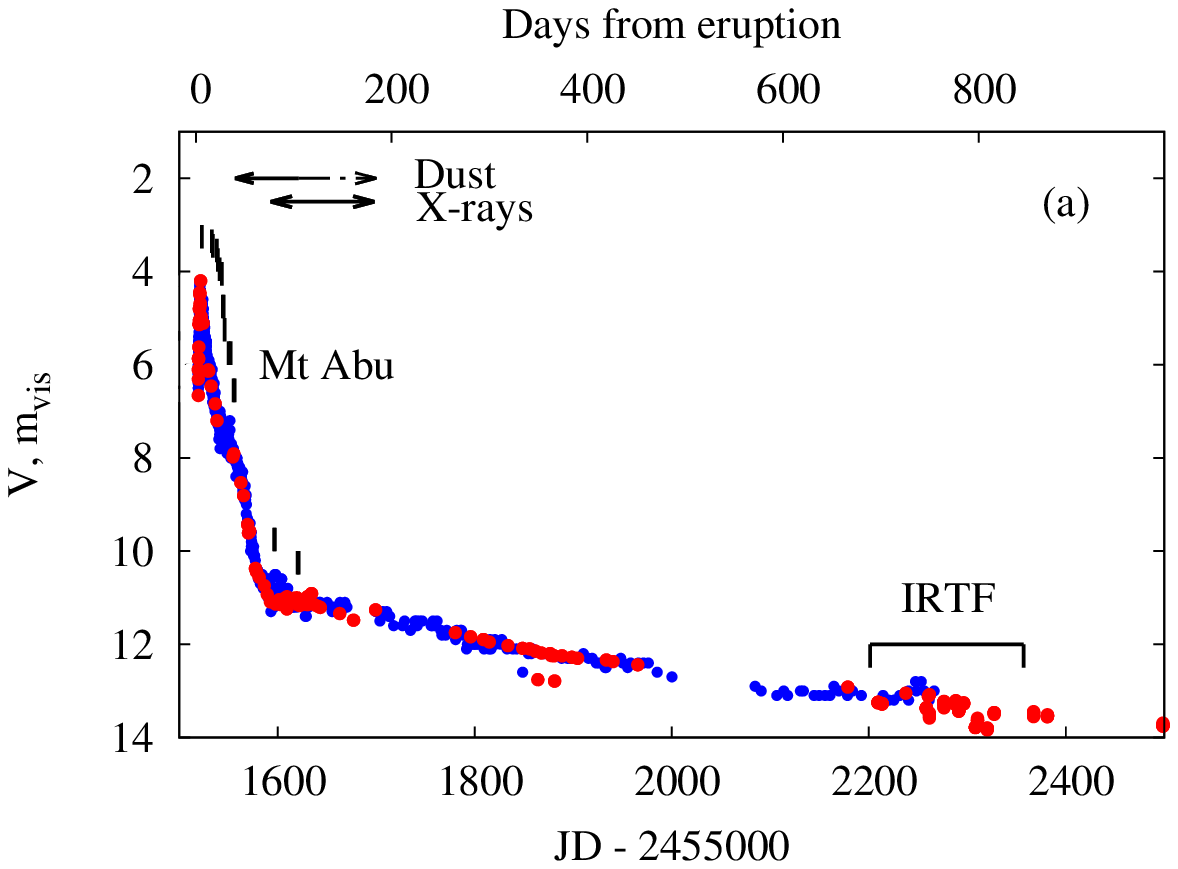}
\includegraphics[angle=0,keepaspectratio,width=\columnwidth]{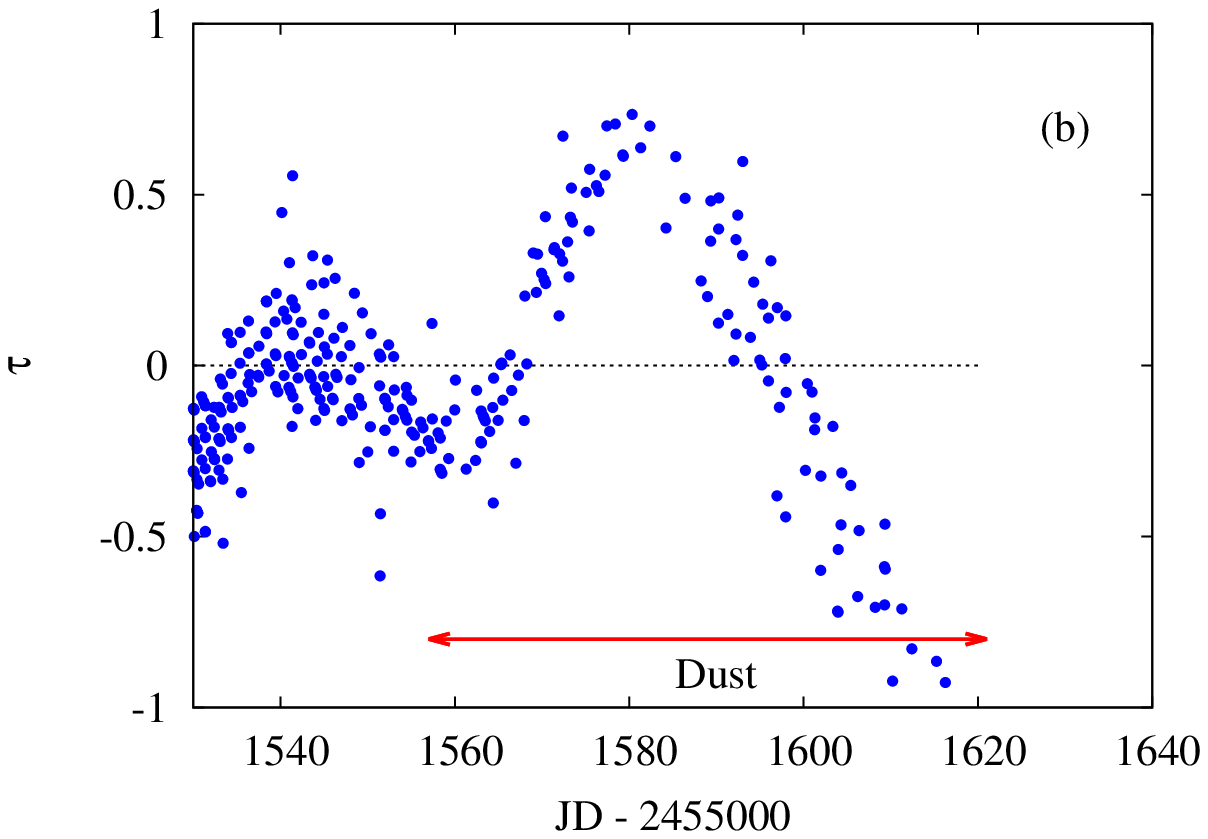}
\caption[]{(a) Visual light curve of \vd.
Blue, visual observations; red, CCD $V$-band observations.
Times of the infrared observations discussed in this paper are indicated by
vertical tick marks, as
are the approximate durations of the dust and X-ray phases; the former is
from infrared observaitons reported in this paper, the latter is the 
duration of the SSS phase from \cite{shore2}. 
(b) Visual light curve, rectified by linear decline immediately before the onset of
dust formation, showing dust optical depth. The extent of the dust phase is from (a).
The apparent decline in $\tau$ for $\mbox{JD}-2\,455\,000\gtsimeq1\,600$ is due to the
increasing contribution of emission lines to the visual light. Note that the
horizontal scales are diffent in (a) and (b).
See text for discussion.
Light curve data courtesy of the BAA and The Astronomer magazine.
\label{LC}} 
\end{center}
\end{figure}

\section{V339 Delphini}
\label{v339}
\vd\ (PNV J20233073+2046041) was discovered -- as a naked-eye nova --
by Koichi Itagaki on 2013 August 14.584~UT
\citep[JD 2456519.084;][]{nakano}, and spectroscopically confirmed by \cite{darnley}.
Pre-outburst optical photometry 
showed that it was varying by a few tenths of a magnitude, around $\sim17.5$~magnitude,
up to about 10~days before eruption
\citep{deacon}. Extensive broadband optical photometry following the eruption 
was given by \cite{munari-b}. The visual light curve is given in Fig.~\ref{LC}(a).
\cite{munari-a} determined the interstellar reddening
to \vd\ to be $E(B-V)=0.18$, which we adopt here. 

\cite{schaefer} obtained near-IR interferometry of \vd, measuring its angular size
within a day of the eruption. They observed the expansion of the
remnant and, in conjunction with an assumed ejection velocity
of $613\pm79$\vunit, deduced a
distance of $4.54\pm0.59$~kpc for the nova. They further detected
an ellipticity in the light distribution, suggesting a prolate 
or bipolar structure that may have developed as early as day~2.
On the basis of photometry of the expanding fireball following outburst, 
\cite{gehrz-v339} determined that the distance of \vd\ is $4.5\pm0.8$~kpc,
and that the eruption occurred on 2013 August 13.9 UT (JD 2,456,518.4); in
this paper we adopt this as $t=0$ and $D=4.5$~kpc.

Extensive optical spectroscopy during the early phase was described by \cite{skopal}.
They determined that the effective temperature of the stellar remnant
was in the range 6000--12000~K, and that the stellar remnant was super-Eddington. 
They determined the ejected mass to be a few $\times10^{-4}$\Msun. \citeauthor{skopal}
also reported the emergence of Raman-scattered \pion{O}{vi} $\lambda=1032, 1038$\AA\
at around 6825\AA\ \citep[but see][for a more plausible
counter-interpretation in terms of emission by \pion{C}{i}]{shore}.
\cite{shore2} have undertaken a multi-wavelength study of \vd, including X-ray 
data obtained with {\it Swift}, UV with the {\it Hubble Space Telescope}, 
and ground-based optical observations. The approximate duration of the X-ray phase, 
such that the X-ray count rate is at least 10\% of the maximum count rate 
(see Figure~1 of \citeauthor{shore2}), is indicated in Fig.~\ref{LC}.

\begin{table}
\begin{center}
 \begin{minipage}{140mm}
  \caption{IR spectroscopic observing log for \vd. \label{log}}
  \begin{tabular}{cccccc}
  \hline  
Facility &    Date UT   &  2013/15   &     JD $-$  &   $t$ &  Bands \\
        & YYYY-MM-DD     &   UT Day   &    2450000   & (days)& covered\\\hline
Mt Abu & 2013-08-18 & Aug 18.63 &  6523.13  & 4.73 & $J\!H$ \\
Mt Abu & 2013-08-28 & Aug 28.66 &  6533.16  & 14.76& $J\!H\!K$\\
Mt Abu & 2013-08-29 & Aug 29.69 &  6534.19  & 15.79& $J\!H\!K$\\
Mt Abu & 2013-09-02 & Sep 02.68 &  6538.18  & 19.78& $J\!H\!K$\\
Mt Abu & 2013-09-03 & Sep 03.63 &  6539.13  & 20.73& $J\!K$\\
Mt Abu & 2013-09-05 & Sep 05.63 &  6541.13  & 22.73& $J\!H\!K$\\
Mt Abu & 2013-09-07 & Sep 07.62 &  6543.12  & 24.72& $J\!H\!K$\\
Mt Abu & 2013-09-08 & Sep 08.65 &  6544.15  & 25.75& $J\!H\!K$\\
Mt Abu & 2013-09-09 & Sep 09.60 &  6545.10  & 26.70& $J\!H\!K$\\
Mt Abu & 2013-09-10 & Sep 10.66 &  6546.16  & 27.76& $J\!H\!K$\\
Mt Abu & 2013-09-14 & Sep 14.72 &  6550.22  & 31.82& $J\!H\!K$\\
Mt Abu & 2013-09-16 & Sep 16.65 &  6552.15  & 33.75& $J\!H\!K$\\
Mt Abu & 2013-09-19 & Sep 19.72 &  6555.22  & 36.82& $J\!H\!K$\\
Mt Abu & 2013-09-20 & Sep 20.59 &  6556.09  & 37.69& $J\!H\!K$\\
Mt Abu & 2013-10-30 & Oct 30.69 &  6596.19  & 77.79& $J\!H\!K$\\
Mt Abu & 2013-10-31 & Oct 31.61 &  6597.11  & 78.71& $J$\\
Mt Abu & 2013-11-23 & Nov 23.62 &  6620.12  & 101.72& $J$\\
Mt Abu & 2013-11-24 & Nov 24.56 &  6621.06  & 102.66& $J\!H\!K$\\
IRTF   & 2015-06-28 & Jun 28.41 &  7201.91  & 683.06& $I\!J\!H\!K$\\ 
IRTF   & 2015-12-01 & Dec 1.21  &  7357.70  & 839.30& $I\!J\!H\!K$\\ 
\hline
  \end{tabular}
  \end{minipage}
  \end{center}
  \end{table}

\cite{tarasova} presented optical spectra of \vd\ at resolution $\sim1000$.
On the basis of H$\alpha$ line profiles they concluded that the ejected material
has a disc-polar structure, with the orbital plane of the binary inclined at $\sim65^\circ$.
They also estimated some elemental abundances and find that helium,
neon and iron are close to solar, while nitrogen and oxygen are overabundant relative to solar 
by factors of $120\pm60$ and $8\pm1.6$ respectively. The mass of the ejecta,
over the period 253 -- 382 days after visual maximum, was estimated to be $\sim7\times10^{-5}$\Msun.

\vd\ was observed with {\it Swift} \citep{kuulkers,page-a}
shortly after outburst; no X-ray 
source was detected at the position of the nova which, however, was
detected in all three  ultra-violet (UV) filters of the UVOT
instrument. The nova was weakly detected with {\it Swift} 30~days after the
eruption \citep{page2}; the emission at this time was consistent with shocked gas
in the expanding shell, with no evidence for the super-soft source (SSS) X-ray emission
commonly seen in CNe \citep[see e.g.][]{krautter}. SSS emission was later detected,
55~days after the eruption \citep{page-b}. Quasi-Periodic Oscillations in 
the X-ray emission were reported by \cite{beardmore} and \cite{ness}.
A summary of the X-ray evolution is given by \cite{shore2}.

\begin{figure*}
\setlength{\unitlength}{1mm}
\begin{center}
\leavevmode
\includegraphics[angle=0,width=14.5cm,keepaspectratio]{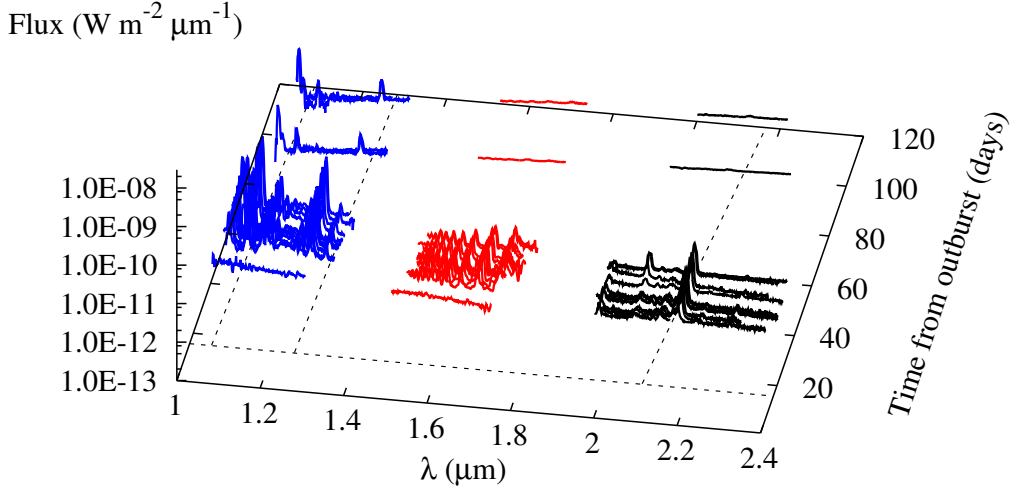}
\caption[]{Evolution of the $J\!H\!K$ spectrum as seen from Mt Abu.
Blue curves: $J$-band data; red curves: $H$-band data; black curves: $K$-band data.
Dashed lines delineate \pion{He}{i} 1.0833\mic,  Pa$\beta$ 1.2821\mic, Br$\gamma$ 2.1662\mic.
The transition from gas emission, with the flux declining to longer wavelengths ($t<40$~days), to gas+dust
emission, with flux rising to longer wavelengths ($t>77$~days), is evident.
The dashed line at $t=15.79$~days indicates an approximate baseline for the 15.79~day data,
to give an indication of the flux scale.
\label{ALL}} 
\end{center}
\end{figure*}

\vd\ is one of an  increasing number of CNe that have been detected
as $\gamma$-ray sources \citep{hays,ackermann,ahnen}; the peak in the
$\gamma$-ray flux in \vd\ occurred about 6~days after optical maximum.
In general, the $\gamma$-ray emission is thought to 
originate from
the interaction between the nova ejecta with a pre-existing red-giant
wind \citep[e.g.][]{tatischeff, martin}. However, \vd\ is one
of only a handful of novae detected at $\gamma$-ray energies in which the
donor is a main-sequence star, and therefore
a different interpretation for the origin of the $\gamma$-ray emission
is required \citep{ackermann}. The first $\gamma$-ray detected nova
with a main-sequence star, V595 Monocerotis, provided a vital clue
in the bipolar morphology of the ejecta \citep{shore13, ribeiro,linford}.
The bipolar morphology was interpreted as originating from the interaction
of the ejecta with the motion of the binary system, allowing
gas to be expelled freely in the polar directions, while within the
equatorial plane material flowed more slowly; this set up a system
of shocks, observed as synchrotron emission at radio frequencies,
at the interface between the equatorial and polar regions where the
$\gamma$-ray production was proposed to occur \citep{chomiuk}.

IR photometric observations of \vd\ were reported by \cite{cass-a,cass-b},
\cite{shenavrin}, \cite{taranova} and \cite{gehrz-v339}. 
\cite{taranova} detected an IR excess
due to dust formation approximately
one month after maximum. They estimated a dust temperature and mass of 
$\sim1500$~K and $\sim1.6\times10^{24}$~g~ ($\sim8\times10^{-10}$\Msun) respectively
on 2013 September 21 (day~38.9), and $\sim1200$~K and $\sim10^{25}$~g~
($\sim5\times10^{-6}$\Msun) on 2013 October 11 (day~58.8). The presence of 
dust is confirmed by the IR photometry by \cite{cass-a} and \cite{gehrz-v339}. 
\citeauthor{gehrz-v339} estimated the amount of dust to be $\sim1.3\times10^{-9}$\Msun\
78.66~days after outburst, and $\sim1.2\:\:[\pm0.4]\times10^{-7}$\Msun\ 
102~days after outburst. \cite{skopal} determined that the dust was 
located beyond the neutral hydrogen zone, where it was shielded from the hard
radiation field of the stellar remnant \citep[see e.g.][]{ER,SCW}.

IR spectroscopy of \vd\ was reported by \cite{stringfellow} and \cite{banerjee-a,banerjee-b}.
These authors reported the presence of \pion{H}{i} recombination lines, together 
with \pion{He}{i}, \pion{O}{i} and \pion{C}{i} emission lines, the spectral
evolution being typical of the taxonomic ``\pion{Fe}{ii} class'' of CNe
\citep[see][for a definition of the various CN spectral classes]{williams}. The eruption most
likely occurred on the surface of a CO white dwarf; such ``CO novae'' are often
copious producers of dust. However, first overtone emission by CO, a common
precursor to dust formation in CNe \citep[e.g.][]{evans-CO,rudy,das,raj,banerjee-16},
was not detected in any of these observations. It would have been detected had it
been present to the extent seen in other novae; its weakness in \vd\ may be
connected with this nova's inability to form a copious amount of dust.

\cite{gehrz-v339} have described IR observations of \vd\ using
{\it SOFIA} and, using their fireball-derived distance, deduced
an outburst luminosity of $\sim8.3\times10^5$\Lsun: \vd\ seems to have been the 
most luminous CO nova on record. They determined the mass of ejected gas to
be $\sim7.5\times10^{-5}$\Msun, and that the gas-to-dust ratio in \vd\ was
in the range $\sim470-940$. This implies that dust formation in \vd\ was much less efficient
than is the case for other CO novae. The inefficient dust
formation in \vd\ may be connected to the weak or absent CO first overtone emission.

\vspace{-5mm}

\begin{figure*}
\begin{center}
\leavevmode
\includegraphics[angle=0,keepaspectratio,width=8cm]{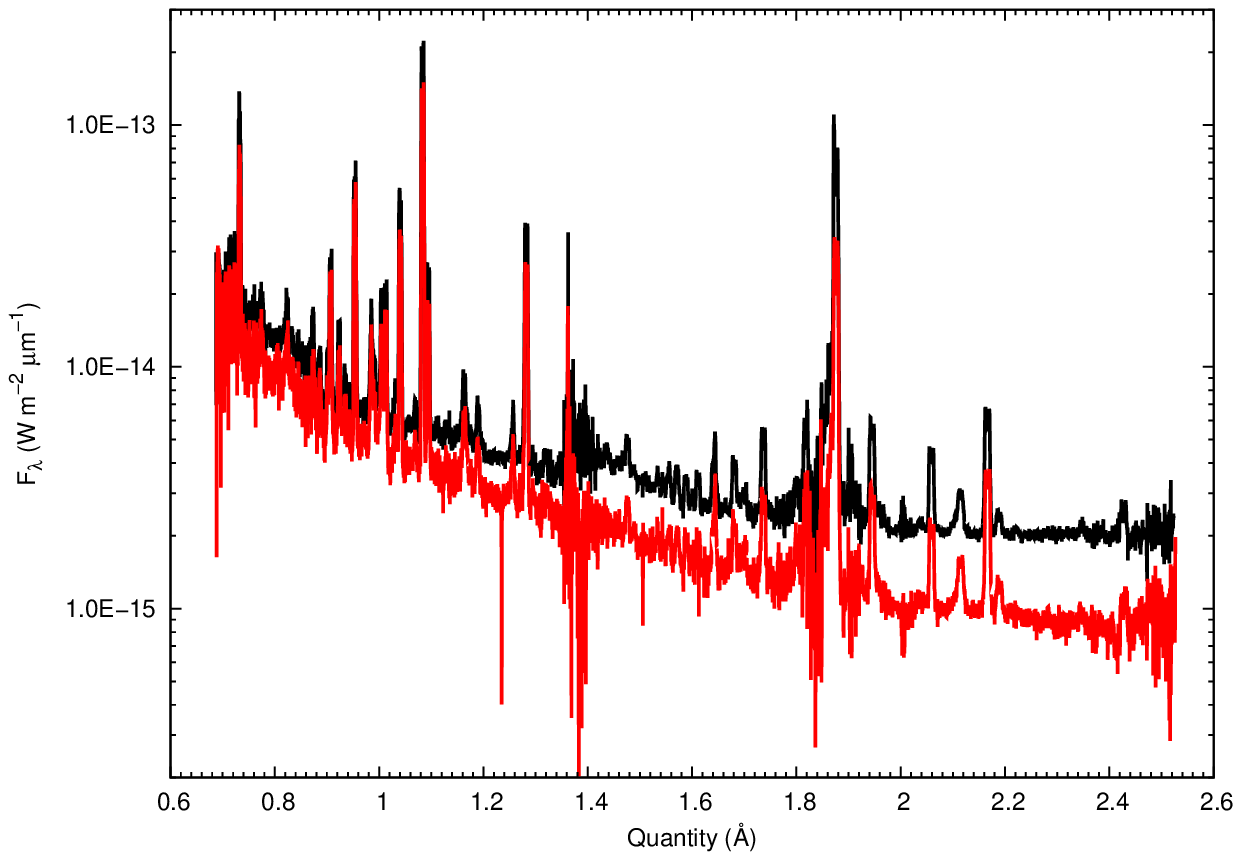}
\includegraphics[angle=0,keepaspectratio,width=8.5cm]{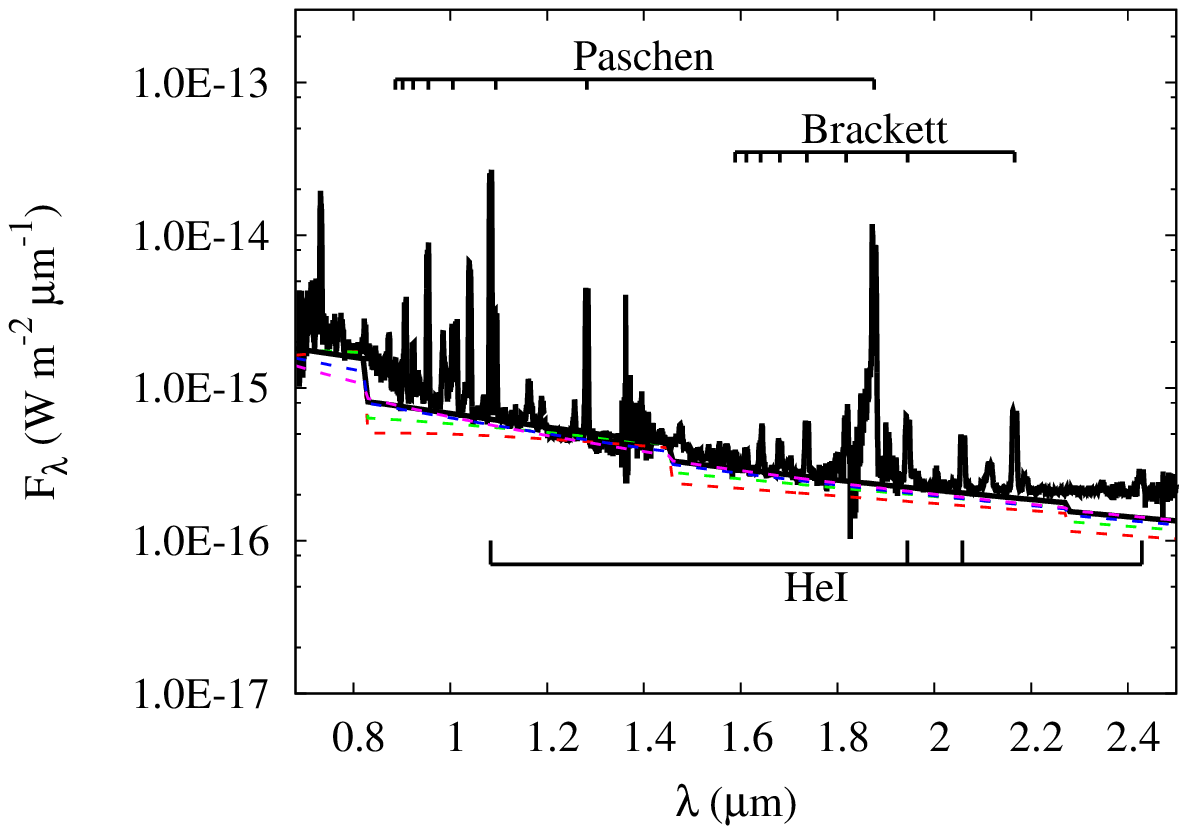}
\caption[]{Left: IR spectra of \vd\ obtained at the IRTF; black spectrum obtained
on 2015 Jun 28.46 (day 683.06), red spectrum on 2015 Dec 1.21 (day 839.30). 
Right: Spectrum obtained on the IRTF on 2015 June 28.46, with prominent emission lines identified.
The dotted curves are free-free and free-bound emission by a gas having electron temperature
$4\times10^4$~K (magenta), $2\times10^4$~K (blue), $10^4$~K (green) and $8\times10^3$~K (red).
The solid black curve is the best fit for free-free emission with electron temperature $1.5\times10^4$~K.
Note the weak dust excess at wavelengths $\gtsimeq2$\mic.
See text for details.
\label{IRTF1}} 
\end{center}
\end{figure*}

\section{Observations}

\begin{figure}
\includegraphics[angle=0,keepaspectratio,width=9cm]{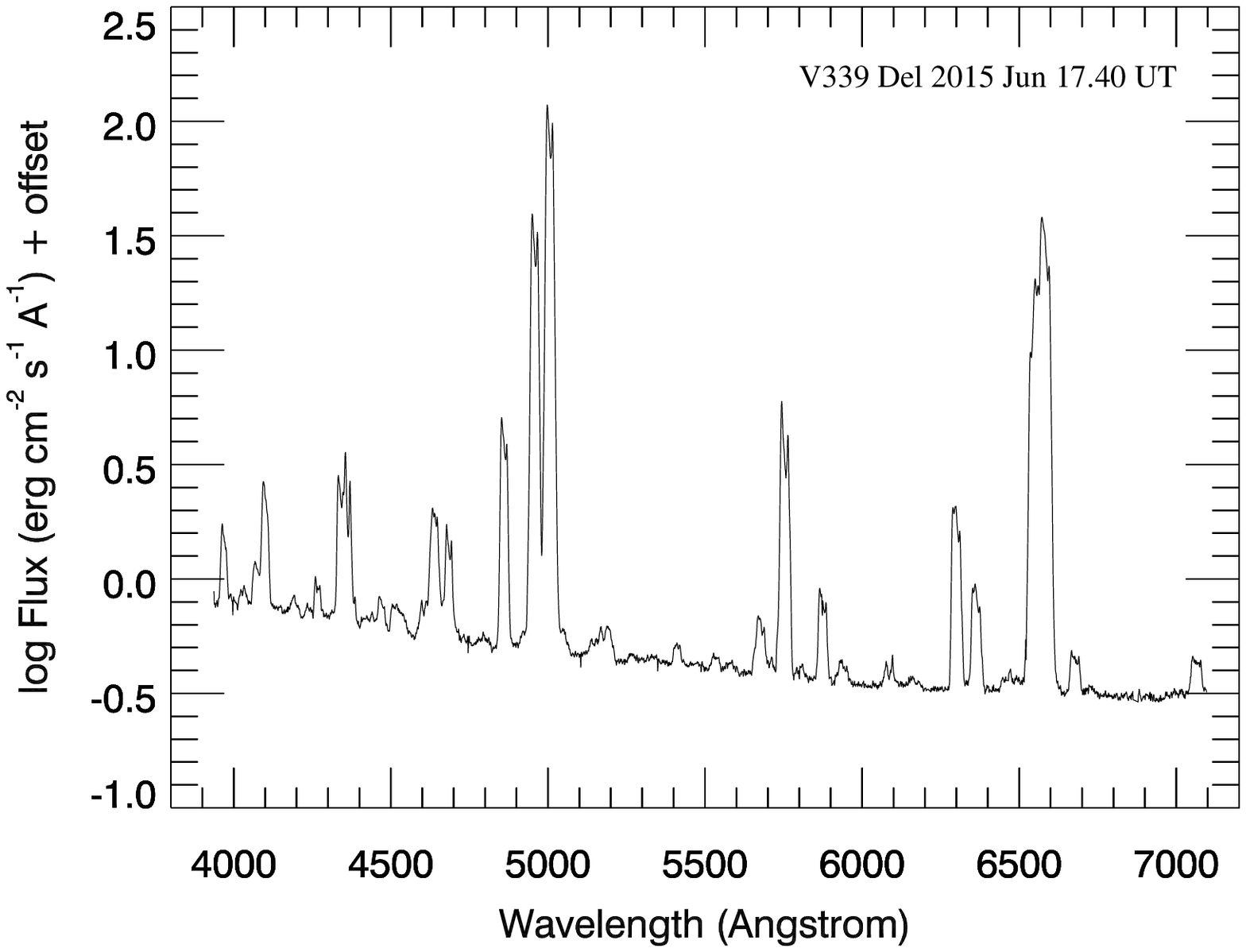}
\caption{Optical spectrum of \vd\ obtained at the MMT Observatory (2015 June 17.40 UT
= day 672.50). The ordinate is log of the observed flux plus an
offset ($=14.0$) to reveal details of the individual line profiles.
The spectrum of \vd\ at this epoch is similar to
other classical novae in the nebular phase \citep{2012AJ....144...98W}, 
exhibiting emission lines of the Balmer series of H, \pion{He}{i} and 
\pion{He}{ii}, \fion{N}{ii}, \pion{C}{ii}, \fion{O}{i}
and \fion{O}{iii}, coinciding with the decline of emission
from various Fe multiplet transitions. The emission line profiles of both H and
metal ions exhibited a ``double-horned'' structure across the line peaks. A complete
identification of spectral lines present in the this spectrum are detailed
in Table~\ref{line_ids_opt} in Appendix~\ref{APP}.} 
\label{fig:mmt_optical_spec}
\end{figure}

IR spectroscopy of \vd\ was obtained at Mt Abu and at the 
3\,m NASA Infrared Telescope Facility \citep[IRTF;][]{becklin};
a summary of the observations obtained is given in Table~\ref{log}, in which $t$ is the time
from the nova explosion reckoned from the $t=0$ date given in Section~\ref{v339}.

\subsection{Mount Abu}

IR photometry in the $J\!H\!K$ bands, and $1-2.5$\mic\ spectroscopy,
of \vd\ were obtained with the
1.2\,m telescope of the Mt~Abu Infrared Observatory \citep[see e.g.][]{BA-BASI} over the period 2013
August 18 -- 2013 November 11.
IR photometry of \vd\ obtained at Mt Abu is included in Table~4 of
\cite{gehrz-v339}, which also gives full details of the observing and data
reduction procedures; the photometric
data and these details are not repeated here.
The spectra are shown in Fig.~\ref{ALL}.

\vspace{-3mm}

\subsection{IRTF}
\label{IRTF2}
Two spectra, covering the 0.78 -- 2.5\mic\ region, were obtained on 2015 June 28.409~UT and 
2015 December 1.20~UT using the 3\,m NASA IRTF.
The integration times for these spectra were 1497~s and 2395~s respectively. The  spectra were 
obtained using SpeX \citep{rayner} in the cross-dispersed mode using the $0.5''\times15''$ slit  
at a resolution of $R = 2000$. Data reduction and calibration was done  using the {\sc Spextool} 
software \citep{cushing} with corrections for telluric absorption being performed 
using the IDL tool {\sc xtellcor} \citep{vacca}.  
The IRTF spectra are shown in Fig.~\ref{IRTF1}.

\begin{figure*} 
\setlength{\unitlength}{1mm}
\begin{center}
\leavevmode
\includegraphics[angle=0,keepaspectratio,width=8cm]{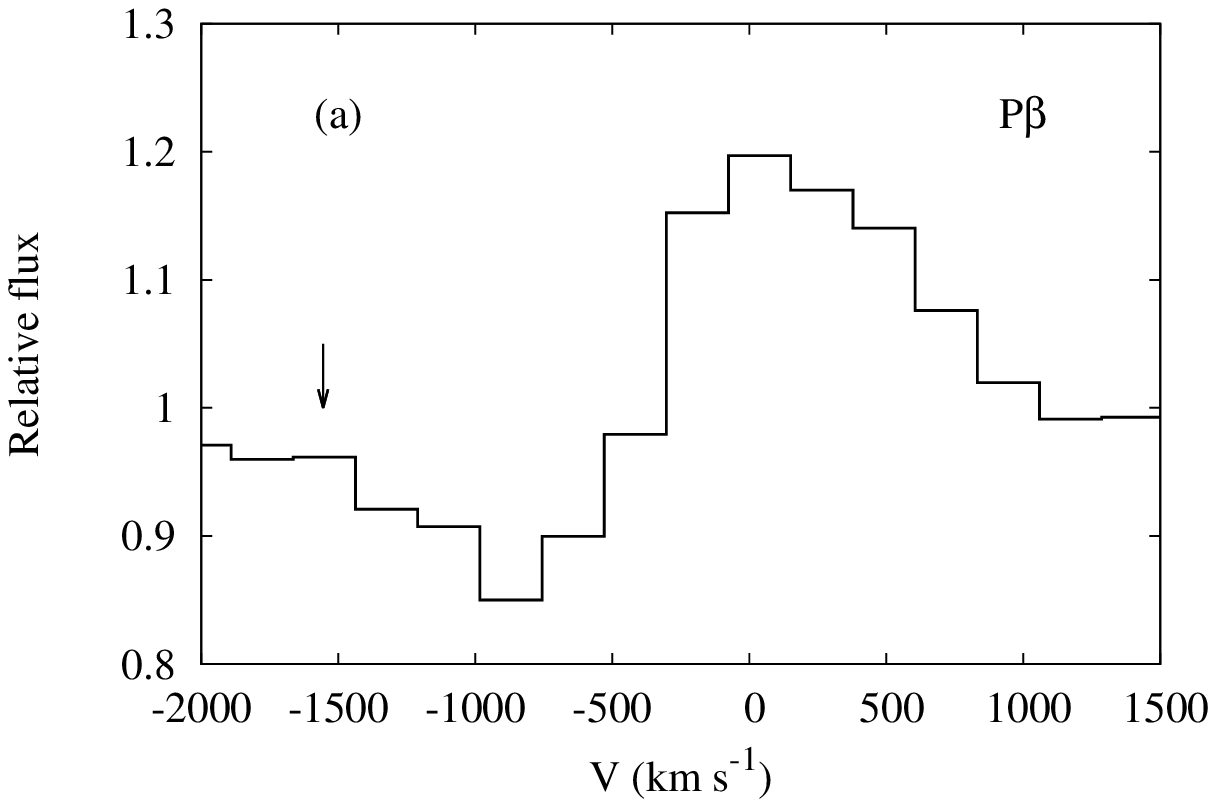}
\includegraphics[angle=0,keepaspectratio,width=8.cm]{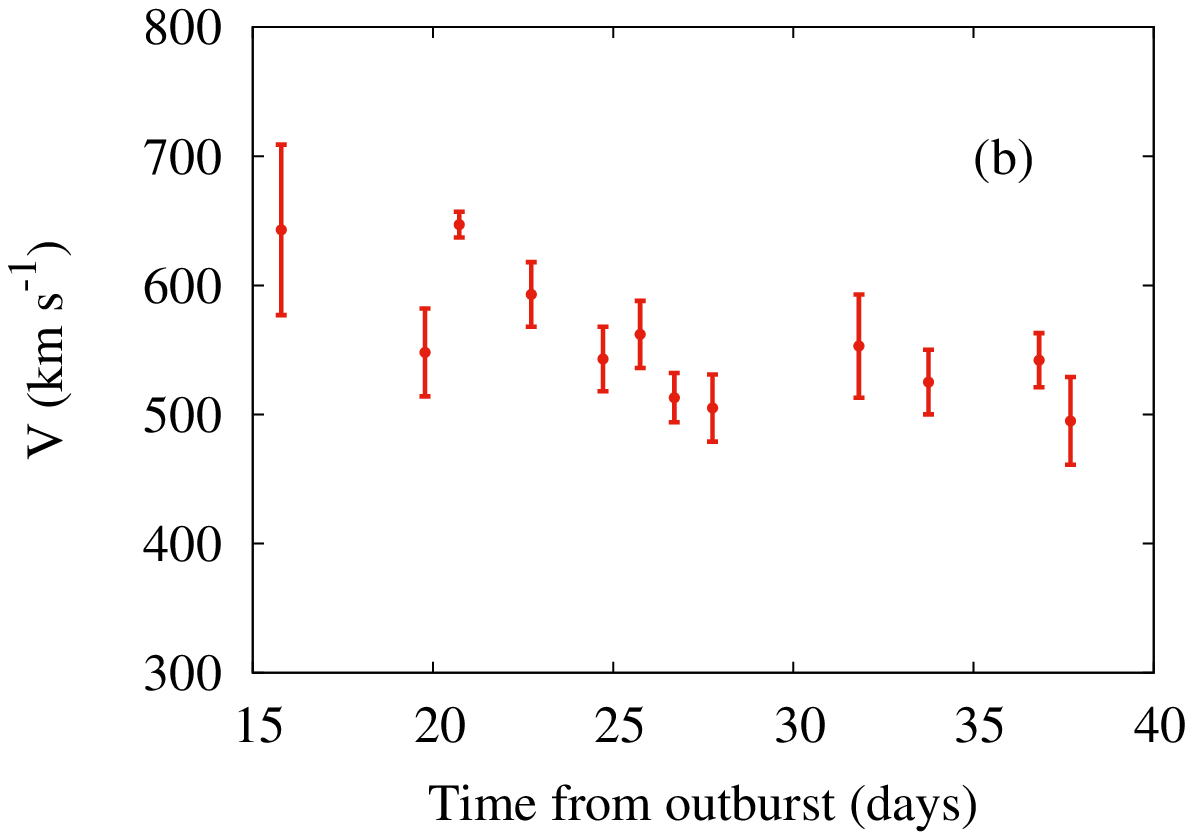}
\includegraphics[angle=-90,keepaspectratio,width=8cm]{Pa-beta_early.eps}
\includegraphics[angle=-90,keepaspectratio,width=8cm]{Pa-beta_late.eps}
\includegraphics[angle=-90,keepaspectratio,width=8cm]{Pa_late.eps}
\includegraphics[angle=270,keepaspectratio,width=8cm]{Vel_plot.eps}
\caption[]{(a) P Cygni profile in Pa$\beta$ on day $t=4.73$~days.
(b)~Expansion velocity as deduced from IR \pion{H}{i} recombination lines over the period 14 -- 37~days.
(c)~Spectrum around 1.282\mic\ Pa$\beta$ line over the period 14 -- 37~days.
(d)~Spectrum around 1.282\mic\ Pa$\beta$ line over the period 77 -- 102~days.
The black horizontal lines in (c) and (d) delineate $\pm500$\vunit.
(e)~Profile of the Pa$\beta$ line on day 77.79;
vertical lines are at the central wavelength of the two components, at 1.2795\mic\
and 1.2839\mic; the corresponding Gaussian profiles (red and blue) and the
overall profile (black) are also shown.
(f) Line profiles in velocity space on day~683.06 for \pion{H}{i} Br$\gamma$ 
(4--7; black), \pion{H}{i} P$\beta$ (3--5; red),
\pion{He}{i}  $^3$S$-^3$P$_o$ (1.083\mic; dark blue)
and \pion{He}{i} $^1$S$-^1$P$_o$ (2.059\mic; light blue). 
\label{VEL}} 
\end{center}
\end{figure*}

The IRTF spectrum obtained on day~683.06 is also shown with line identifications
in Fig.~\ref{IRTF1}. The spectrum shows a number of prominent emission lines, primarily hydrogen
and \pion{He}{ii} recombination lines and \pion{He}{i}.
The emission lines show a castellated structure that is present in both hydrogen and
helium lines. 

In both the Mt Abu and IRTF data the emission lines sit atop a nebular continuum. 
We determine the nebular continuum
using an assumed flux for the Balmer H$\beta$ line, and an electron temperature and density.
We have calculated several nebular continua using {\sc nebcont},
which is part of the STARLINK \citep{currie} {\sc dipso} spectral analysis package \citep{dipso}.
{\sc nebcont}
includes free-bound continua from 
H, He, CNO and Ne, as well as the 2s--1s 2-photon continuum from H and \pion{He}{ii}. 
{\sc dipso} requires electron temperature and density and the H$\beta$ flux (amongst other
parameters) as input. 

For \Ne\ we extrapolate the value in \cite{gehrz-v339} for day 3.25, namely 
$\Ne=10^{13}$~cm$^{-3}$; \citeauthor{gehrz-v339} argue that the electron density declines
with time as $\Ne\propto{t}^{-2}$, giving $\Ne=2.26\times10^8$~cm$^{-3}$ at the time of the
2015 June IRTF observation ($t=683.06$~days).
This is substantially higher than the value ($\sim4\times10^5$~cm$^{-3}$) 
deduced by \cite{shore2} as early as day~435. There are two reasons for this: 
First we note that \citeauthor{shore2}
used the \fion{O}{iii}5007+4959/\fion{O}{iii}4363 flux ratio to determine \Ne; the 5007, 4959\AA\ lines have a 
critical density for collisional de-excitation of $6.8\times10^5$~cm$^{-3}$ at $\Te=10^4$~K \citep{AGN2}.
Since the H recombination line fluxes are $\propto\Ne^2$, the \pion{H}{i} emission must come primarily 
from high density regions, with minimal contribution from low density regions
in which \fion{O}{iii} emission originates. 
Indeed the \fion{O}{iii}5007+4959/\fion{O}{iii}4363 flux ratio 
in Fig.~\ref{fig:mmt_optical_spec}, obtained at about the same time as the 2015 June IRTF
spectrum, is $\simeq67.3$, giving $\Ne \simeq 4.8\times10^5$~cm$^{-3}$, consistent
with \cite{shore2}. And second, we have assumed $\Ne\propto{t}^{-2}$ as opposed to the
$t^{-3}$ dependence assumed by \citeauthor{shore2}.

We explore \Te\ values in the range $0.8-4\times10^4$~K.
To estimate the H$\beta$ flux, a near-contemporaneous optical spectrum of \vd, taken 
by D. Boyd on 2015 July 5, was obtained from the {\it Astronomical Ring for Access to Spectroscopy}
(ARAS) database\footnote{http://www.astrosurf.com/aras/Aras\_DataBase/Novae}.
This was flux calibrated by anchoring the $V$-band centre using $V = 13.3\pm0.2$ magnitudes. 
Using IRAF\footnote{IRAF is distributed by the National Optical Astronomy Observatories, 
which are operated by the Association of Universities for Research in Astronomy, Inc., 
under cooperative agreement with the National Science Foundation.},
the measured H$\beta$ flux was found to be $(2.0\pm0.4)\times10^{-15}$~W~m$^{-2}$;
dereddening by $E(B-V)=0.18$ leads to a dereddened H$\beta$ flux of $\sim3.5\times10^{-15}$~W~m$^2$.
We use this as a guide to fit the nebular continuum and we 
vary the H$\beta$ flux to get a reasonable fit (by eye) to the $J\!H\!K$ data. We used
$\Te=4\times10^4$~K, H$\beta$ flux $1.86\times10^{-15}$~W~m$^{-2}$;
$\Te=2\times10^4$K, $2.82\times10^{-15}$~W~m$^{-2}$;
$\Te=1\times10^4$K, $2.07\times10^{-15}$~W~m$^{-2}$;
$\Te=8\times10^3$K, $2.07\times10^{-15}$~W~m$^{-2}$; and
$\Te=1.5\times10^4$K, $3.63\times10^{-15}$~W~m$^{-2}$. The uncertainties 
in the H$\beta$ fluxes are typically $\pm0.1$~dex.

The observed magnitude of the Paschen discontinuity at the 0.820\mic\ limit, 
and of the Brackett discontinuity at 1.459\mic, show that \Te\ must lie between 
$10^4$~K and  $4\times10^4$~K;
the calculated continuum for $1.5\times10^4$~K seems to give a reasonable fit
to the continuum, but $2\times10^4$~K provides a better fit to the magnitude of the 
discontinuities. Where necessary we assume the latter value in what follows.

It is evident in Fig.~\ref{IRTF1} that there is a weak excess longward of
$\sim2$\mic\ on day 683.06; this is residual emission due to dust, which was
far more prominent
in the early evolution of \vd\ (see Section~\ref{DUST} below).

\subsection{MMT}

An optical spectrum of \vd\ was obtained on
2015 June 17.40 UT (day 672.50) at the 6.5~m MMT with the Blue Channel
Spectrograph \citep{1989PASP..101..713S}.  A 1\arcsec\ $\times$ 180\arcsec\
long--slit was used with a 500 line per mm grating and a thinned STA
$2688 \times 512$ pixel detector covering all or part of the
3800--7100 \AA\ region at a nominal resolution of 3.6 \AA.  A UV--36 long--pass
filter was used to block 2nd order light from contaminating the red portion of
our spectra. Spectra of a HeArNe lamp provided wavelength calibration while
the spectra of a quartz--halogen lamp provided flatfield correction images. 
Twenty, individual  10-s exposure spectra were extracted and coadded to produce the final
spectrum, which is shown in Fig.~\ref{fig:mmt_optical_spec}.  
The spectrophotometric standard star Kopff 27 was obtained to provide            
flux calibration.

The data were reduced using standard IRAF 
packages and standard spectral extraction and calibration 
techniques were used.

\vspace{-3mm}

\section{The emission lines and nebular continuum}

\subsection{Emission lines and line fluxes}
\label{mtabu}

The  spectra of \vd\ in the $1-2.5$\mic\ region, taken from Mt Abu,  
are shown in a compact fashion in Fig.~\ref{ALL};
an expanded view of representative spectra, where individual lines are more clearly visible,
is  shown in Fig.~\ref{dust-fig}, which is further discussed below.  
Among the emission  lines seen during the early phase
are prominent lines of hydrogen from the Paschen and Brackett series 
(viz. Pa 5--3, 6--3, Br 7--4 and Br 10--4  through  Br 17--4).
Helium lines are weak during this stage but increase in strength later;
the main lines seen are \pion{He}{i} 1.0833\mic, 2.0581\mic. \pion{N}{i} lines are few,
the strongest \pion{N}{i} feature being the 1.2461/1.2469\mic\ line. 
In addition, there is  a cluster of weaker \pion{N}{i} features, 
blended with many \pion{C}{i} lines,  lying between 1.2 and 1.275\mic.  
The most prominent \pion{O}{i} lines are the Ly$\beta$ fluoresced 1.1287\mic\ line,  
which is one of the strongest lines in the spectrum,  and the relatively weaker, 
continuum excited \pion{O}{i} 1.3164\mic\ line.   
A large number of prominent carbon lines are seen which include 
\pion{C}{i} lines at 1.165, 1.175, 1.188, 1.689\mic, as well as the forest of 
strong \pion{C}{i} lines between 1.74---1.8\mic\ at the $H$-band edge. 
As discussed in \cite{BA-BASI}, these \pion{C}{i} lines  
are the easiest way to demarcate the \pion{Fe}{ii} from the He/N class of novae in the near-IR.  
In brief, the near-IR spectra of \vd\ are typical of the 
IR properties of the \pion{Fe}{ii} class of novae. 
Several examples of the near-IR spectra of novae of this class, such as V1280~Sco, V2615~Oph, V476~Sct, etc.,
are given in \citeauthor{BA-BASI} and references therein.  
A detailed identification of the lines seen in the spectra of \vd, along with line fluxes 
-- uncorrected for extinction -- are listed in Tables~\ref{line_ids_e} and
\ref{line_ids_l} in Appendix~\ref{APP}.

\subsection{Expansion velocities}

On day 4.73 there is a clear P~Cygni profile in the Pa$\beta$ line (see Fig.~\ref{VEL}(a)),
with a suggested terminal velocity of $\sim-1\,500$\vunit; a
P~Cygni profile with similar terminal velocity is also present in the \pion{C}{i} 
$^3$P$_2-^3$P$_2^o$ line at 1.2717\mic. These were no longer present by day~14.76. 
In the optical, \cite{skopal} found that the
H$\alpha$ line displayed a P-Cyg profile, with terminal velocity
$-1\,600$\vunit\ at $t\simeq0.95$~day, decreasing to $-730$\vunit\ at
$t\simeq6$~days. Despite the lower resolution of our IR data at this time,
the IR terminal velocity seems consistent with that in the optical.

The Half Width Half Maximum (HWHM) of the emission lines can be used to
estimate the expansion velocity $V$ of the ejecta. 
While the Full Width at Zero Intensity (FWZI) captures high velocity wings
(see Fig.~\ref{VEL}(c),(d),(f) below) the HWHM better characterises the bulk
of the ejecta. In Fig.~\ref{VEL}(b) we show
the dependence of the HWHM velocities (corrected for instrumental resolution)
on time $t$ since outburst. There may be a decline in $V$ over the period
$t=14-37$~days (see Fig.~\ref{VEL}(b)) but this is marginal at best.
The weighted mean of the velocites in Fig.~\ref{VEL}(b) is $583$\vunit\ and
we assume this value in what follows.

From $t\simeq77$~days, however, there
seems to have been a significant change in the profile of the Pa$\beta$ line
(Fig.~\ref{VEL}(c)-(d)). Prior to day~37 Pa$\beta$ is symmetrical, with HWHM $\simeq530$\vunit\ but by day~77,
there is a distinct asymmetry. A simple Gaussian fit to the Pa$\beta$ line for day~77.79 shows evidence for
two distinct features, with HWHM velocites of 418\vunit\ (blue component) and 437\vunit\ (red component),
deconvolved for the instrumental resolution.
The line centroids are at velocities of $-622$\vunit\ and $+407$\vunit, respectively (Fig.~\ref{VEL}(e)).
This change in profile coindides with the epoch of maximum dust mass (see Section~\ref{DUST}).
The data presented here suggest that the torus was obscuring the receding lobe as
early as day~77. By day 683.06, however, the blue wing seems to have recovered
(see Fig.~\ref{VEL}(f)), possibly due to
dispersal of the dust or, more likely, due to the destruction of a large fraction of the dust
shell (see Section~\ref{dest} below).

\subsection{Case B analysis}

\begin{figure}
\begin{center}
\leavevmode
\includegraphics[angle=0, bb =  142 128 534 428, width = 3.0in, clip]{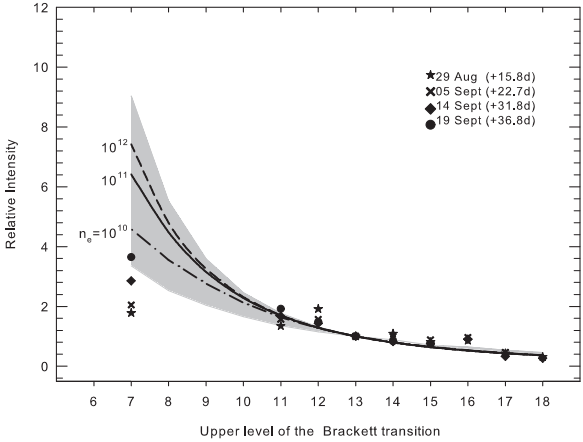}
\caption[]{The relative intensities of the Brackett lines, with Br-13 set to unity,
for four early epochs of the Mt Abu observations (viz. 2013 Aug
29, Sept 5, 15 and 19). The parameter space for the predicted
Case~B values is shown in grey for combinations of electron density and
temperature varying between $10^5$ to $10^{14}$~cm$^{-3}$, and $\Te=5\,000$ to 20\,000~K
respectively. Three specific Case~B curves are shown, for $T = 10\,000$~K and
$\Ne = 10^{10}, 10^{11}$ and $10^{12}$~cm$^{-3}$ (dash-dot, dashed and 
continuous curves respectively). Such high densities are expected in the 
ejecta during the early evolution.\label{caseB}} 
\end{center}
\end{figure}

A recombination Case~B analysis was carried out for four epochs (Fig.~\ref{caseB}) between 
15.8 to 36.8~d after outburst, before dust formation set in.
The analysis was along similar lines as for Nova Cep 2014
\citep{srivastava}. Fig.\ref{caseB} shows the Brackett line strengths  
with respect to Br13 set to unity. The line fluxes do not match predicted Case~B
values and the Br$\gamma$ line strength is significantly lower than 
the values predicted by
 \cite{storey}. This indicates that Br$\gamma$ is optically thick, and 
so possibly are the other Br lines. Similar optical depth effects in the Br lines, 
due to high plasma densities during the early stages after outburst, 
is common and has been seen in several other novae  e.g. Nova Oph 1998 \citep{lynch}, V2491~Cyg,
V597~Pup, RS~Oph, T~Pyx, Nova Cep 2014 \citep[see][and references therein]{naik, joshi, srivastava}.

\begin{table*}
 \centering
\begin{minipage}{140mm}
  \caption{Dust evolution in \vd. 
 ``BB'' denotes black body grains, ``AC'' denotes amorphous carbon, ``GR'' denotes
 graphitic carbon.
 For the black body case, a grain radius $a=1$\mic\ is assumed. \label{dust-tab}}
  \begin{tabular}{ccccccccc}
  \hline  
 $t$ & $[\lambda\,f_\lambda]_{\rm max}$ &  $T_{\rm d}$ & \multicolumn{3}{c}{$M_{\rm d}$ ($10^{-9}$\Msun)} & &
 \multicolumn{2}{c}{$a (\mu$m)}  \\\cline{4-6} \cline{8-9}
 (days)& ($10^{-12}$ W m$^{-2}$) & (K) & BB & AC & GR &&   AC & GR  \\\hline
 36.82 & $8.41~[\pm0.67]$& $1\,637\pm65$ & $25.6\pm0.58$ & $1.66\pm0.34$ & $2.33\pm0.52$ & &2.97 & 4.29 \\
 37.69 & $8.85~[\pm1.38]$& $1\,365\pm44$ & $55.8\pm1.31$ & $4.15\pm0.91$ & $6.44\pm1.48$  && 6.72& 10.64\\
 77.79 & $17.2~[\pm0.3]$ & $1\,014\pm3$  & $356\pm0.7$ & $33.1\pm0.74$ &$6.07\pm1.42$  &&6.46 & 11.93\\
  102.14& $2.59~[\pm1.02]$& $854\pm74$ & $107\pm5.52$ & $11.28\pm6.40$ & $22.79\pm1.37$  && 8.47& 1.07\\
 102.66& $14.1~[\pm0.4]$ & $1\,017\pm4$  & $289\pm0.93$ & $26.76\pm0.91$ & $49.02\pm1.72$  && 3.66 & 6.74\\
 683.06& $2.93~[\pm{}]\times10^{-2}$  & $684\pm{\sim50}$ & $0.37[\pm0.16]$ & $0.84[\pm0.0.41]$ & $0.87[\pm0.42]$  && 0.54& 1.23 \\ 
\hline
  \end{tabular}
  \end{minipage}
\end{table*}

\begin{figure*}
\begin{center}
\leavevmode
\includegraphics[angle=0,keepaspectratio,width=7.8cm]{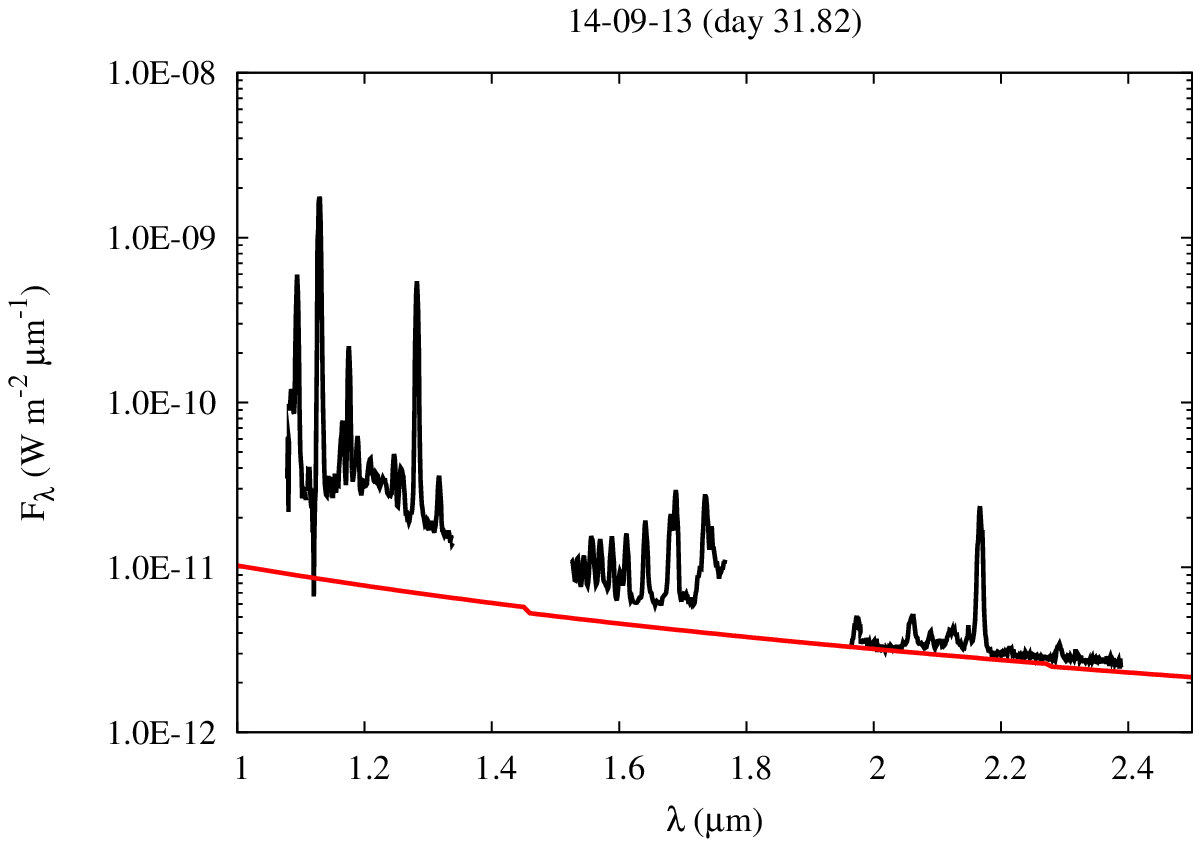}
\includegraphics[angle=0,keepaspectratio,width=7.8cm]{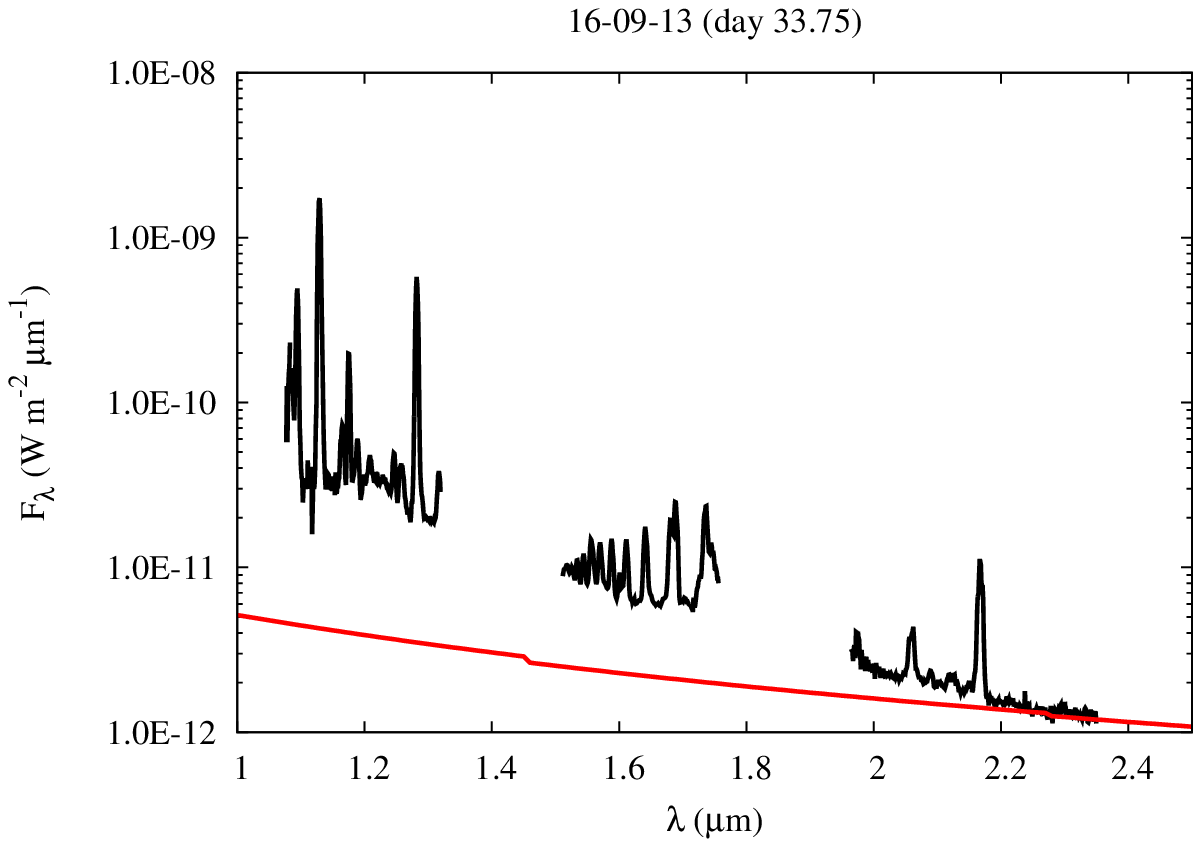}
\includegraphics[angle=0,keepaspectratio,width=7.8cm]{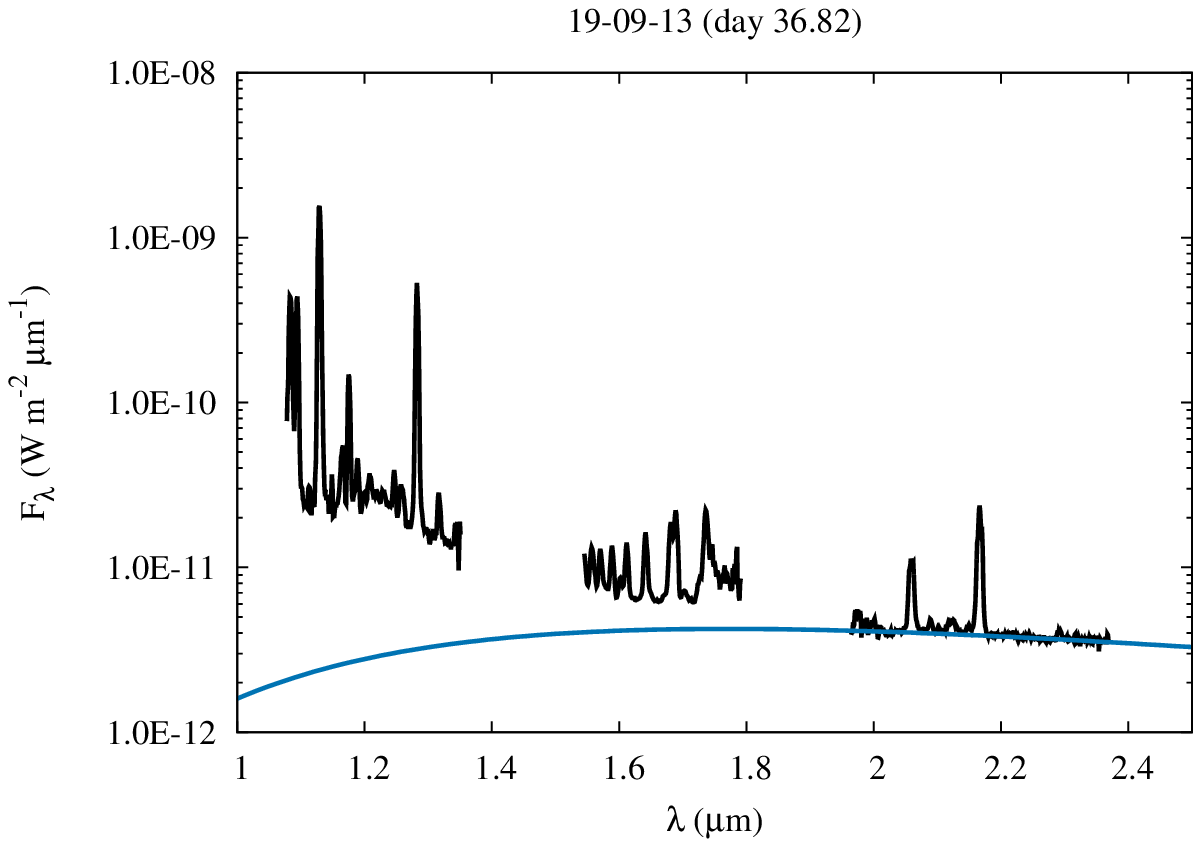}
\includegraphics[angle=0,keepaspectratio,width=7.8cm]{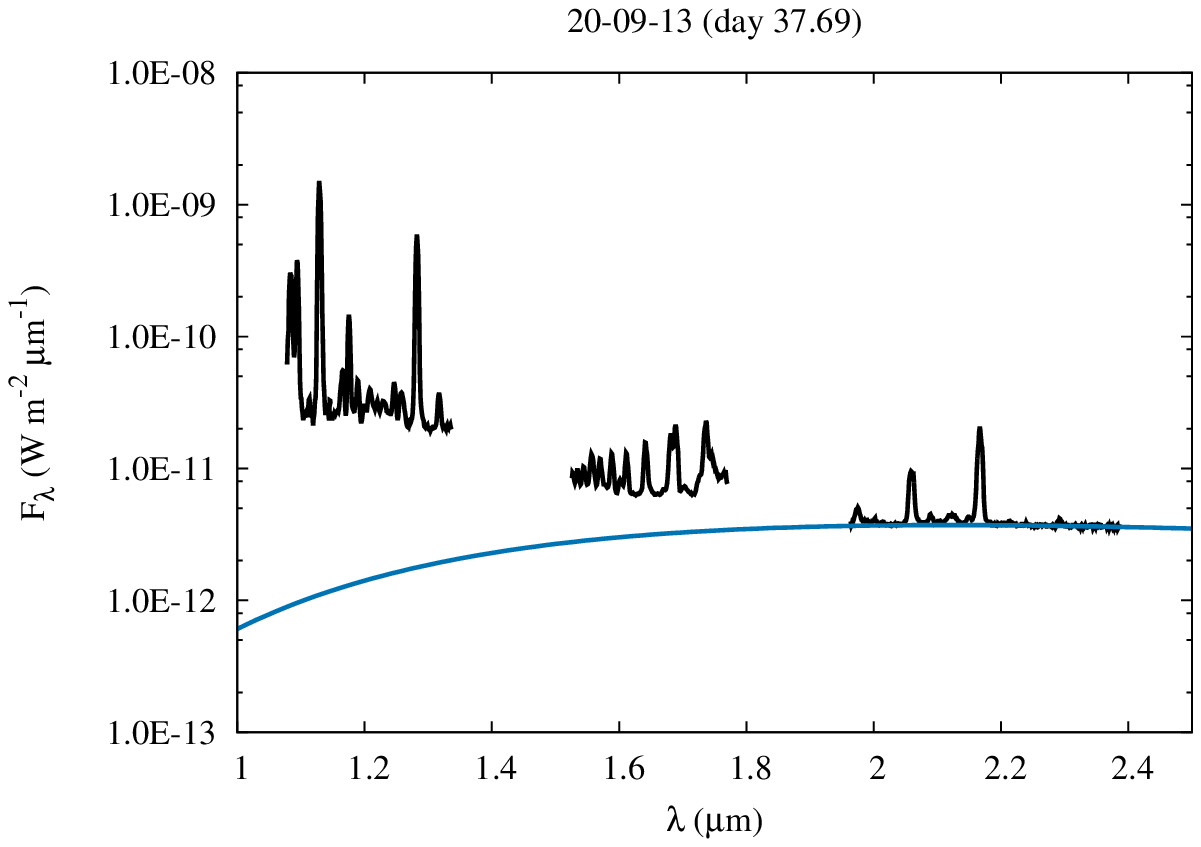}
\includegraphics[angle=0,keepaspectratio,width=7.8cm]{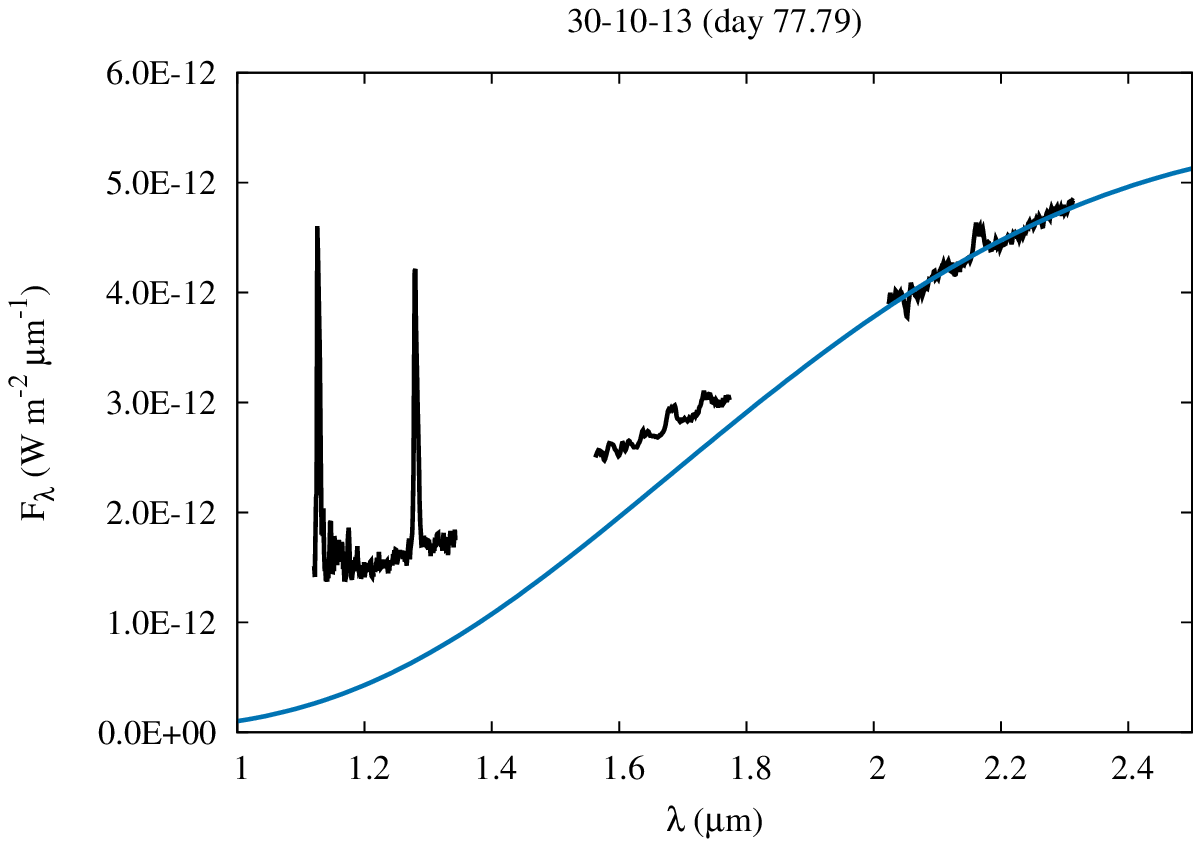}
\includegraphics[angle=0,keepaspectratio,width=7.8cm]{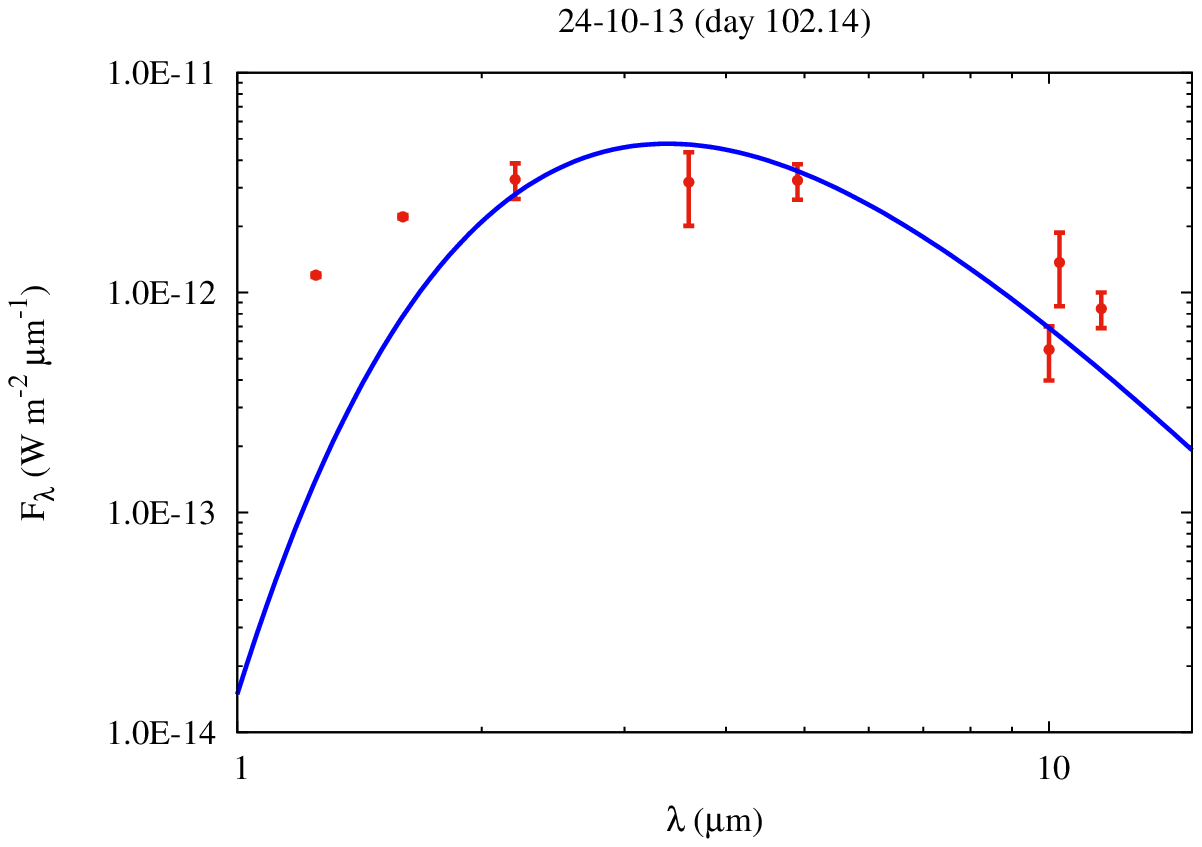}
\includegraphics[angle=0,keepaspectratio,width=7.8cm]{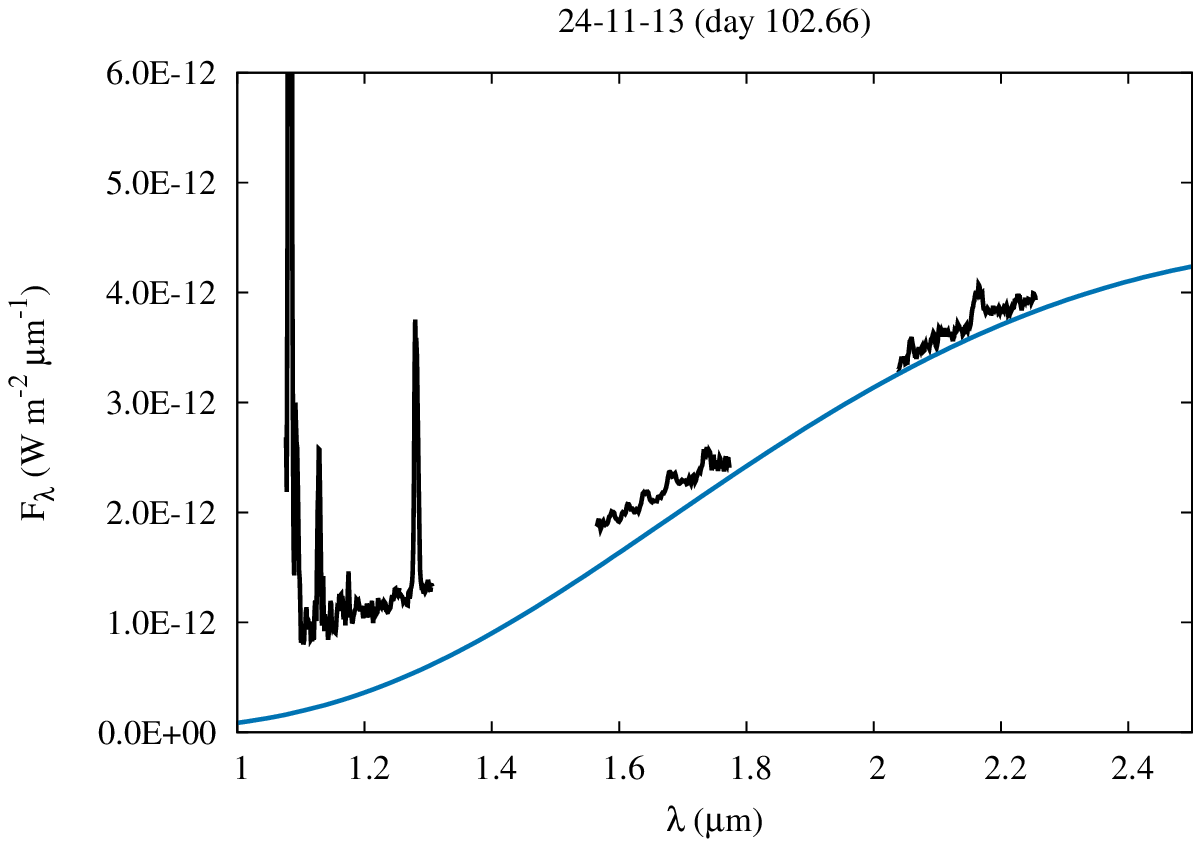}
\includegraphics[angle=0,keepaspectratio,width=7.8cm]{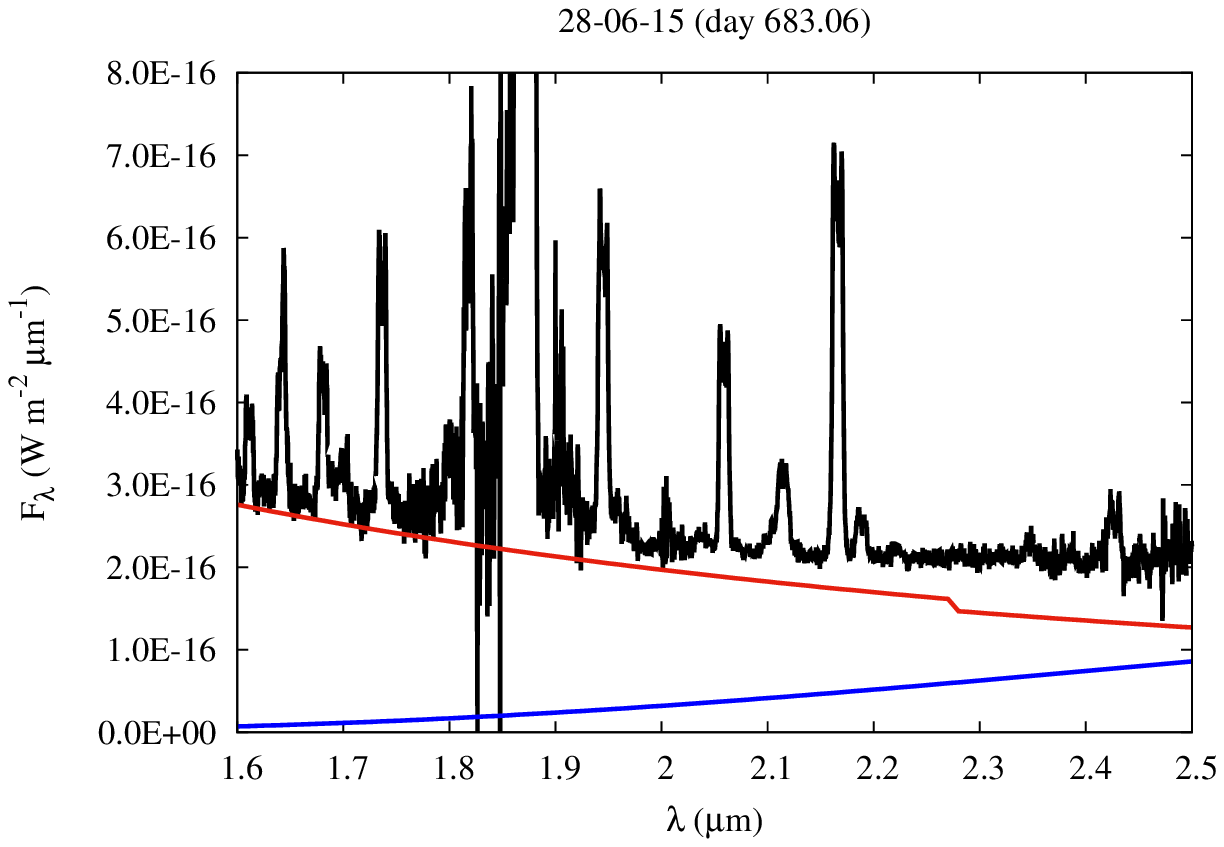}
\caption[]{Evolution of the IR spectral energy distribution from day 31.82 to
day 683.06. 
Data, shown as black lines, dereddened by $E(B-V)=0.18$.
Blue lines, black body function fitted to $K$-band data;
red lines free-free and free-bound emission.
The 2013 November 24 (day 102.14) plot shows the O'Brien Observatory photometry
reported in \cite{gehrz-v339}, dereddened by $E(B-V)=0.18$.
For the 2015 June 28 plot, the additional red curve is the contribution from
free-free and free-bound emission. See text for details.
\label{dust-fig}} 
\end{center}
\end{figure*}

Following the formalism of \cite{hummer} and \cite{storey}, the optical depth at the
Br$\gamma$ line-center is given by 
\[ \tau_{n,n'}  = \Ne\,\Ni \: \Omega(n,n') \: R \:\:, \]
where \Ne\ and \Ni\ are the electron and ion densities respectively, which are assumed to be equal,
and $\Omega(n,n')$ is the  opacity corresponding to the transition from upper level $n$ to lower 
level $n'$, values of which are listed in \cite{storey}. The path length $R$ is taken as 
the kinematical distance $R=Vt$ travelled by the ejecta, where $V = 583$\vunit\ is the velocity of ejecta 
and $t$ is the time after outburst. We consider values of $t$ between 
15.8~day to 36.8~day (from Fig.~\ref{caseB}) and set the constraint that 
$\tau \mbox{(Br$\gamma$)}= \Ne\Ni \Omega(n,n') R$ be greater than 1. 
The lower limit on the electron density \Ne\ is then found to be in the range 
$0.41\times 10^{10}$ cm$^{-3}$ to $0.98 \times 10^{10}$ cm$^{-3}$ on day 15.8. 
By day 36.8, the density decreases in the range  $0.27\times 10^{10}$ 
cm$^{-3}$ to  $0.64 \times 10^{10}$ cm$^{-3}$.  
These derived lower limits should be smaller than the actual \Ne\ values because 
$\tau\mbox{(Br$\gamma$})$ can be considerably $> 1$. An additional caveat 
in this analysis is the  intrinsic assumption of spherical geometry for the ejecta. 
\citep*{slavin} since a bipolar morphology  
appears more appropriate for \vd\ \citep[see also][]{schaefer,shore2}.

A rough estimate of the ejected mass  may be obtained using  
\[M_{\rm ej} = \phi\, V \Ne m_{\rm H} \:\:, \]
where $V( = 4/3\pi{R}^{3}$) is the volume, $\phi$ is the volume filling factor
\citep[assumed $= 0.1$ from][]{shore2} and $m_{\rm H}$ is the proton mass. 
We use the lower limits on \Ne\  estimated above and allow  $R$ to vary between the 
distance traversed from 15.8~d to 36.8~d. The lower limit on the mass $M_{\rm ej}$ 
is estimated to lie between $0.05 \times 10^{-5}$ to $0.17 \times 10^{-5}$\Msun. 
These estimates are roughly a 
factor of 10 -- 20 times lower than the mass estimates given by \cite{gehrz-v339} and \cite{shore2}, 
which lie in the range  $(1 - 3) \times 10^{-5}$\Msun, but as noted
above our values are lower limits.

\begin{figure}
\includegraphics[angle=0,keepaspectratio,width=7.8cm]{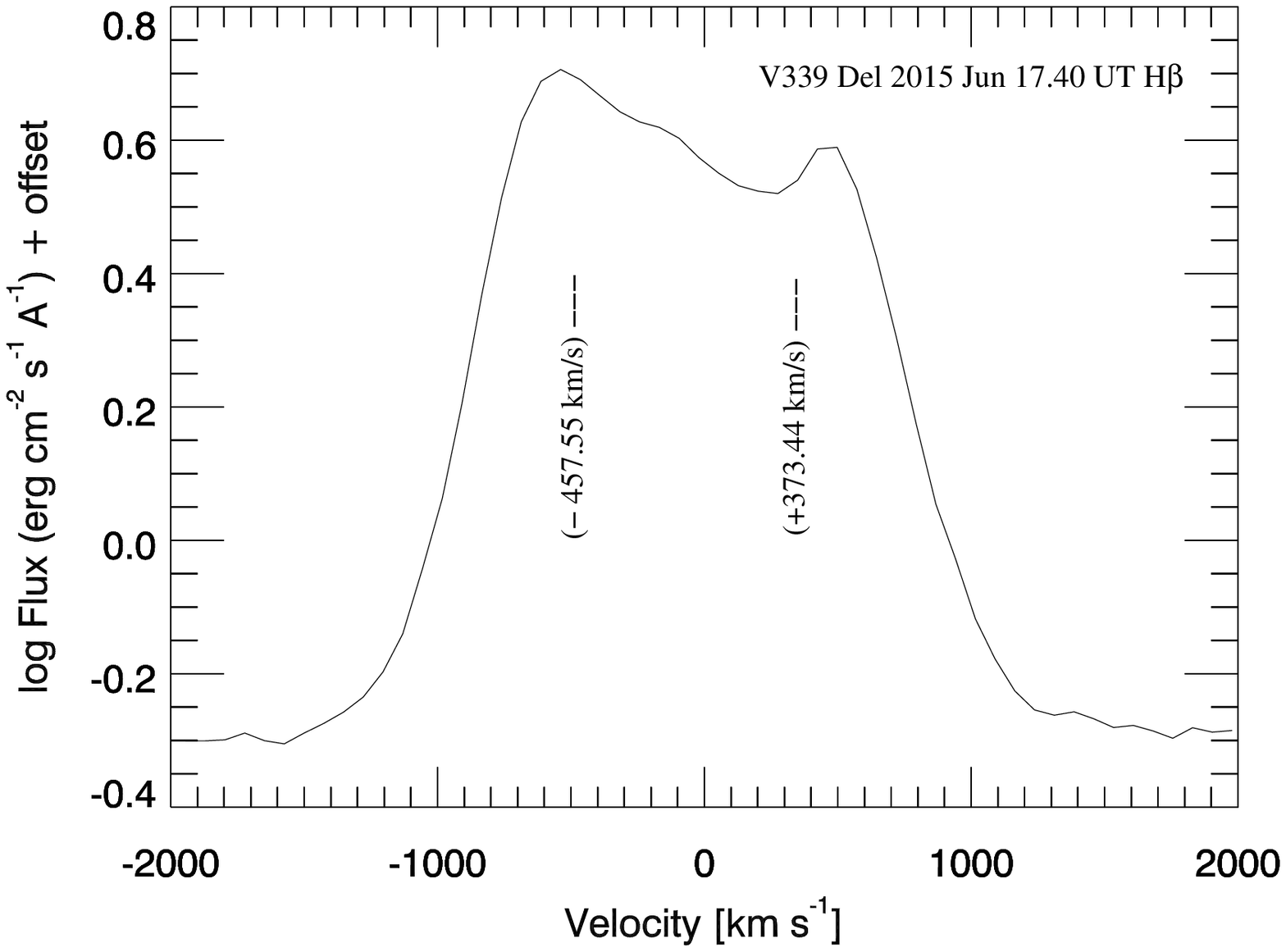}
\caption{The H$\beta$ line profile of \vd\ observed on 2015 June 17.40 UT, extracted from the
spectra shown in Figure~\ref{fig:mmt_optical_spec}. Fitting a simple two-Gaussian
model to deblend the velocity components of the H$\beta$ profile reveals two
peaks, one blue shifted by 457.55\vunit\ and one red shifted by 373.4\vunit\ 
from the rest wavelength of 4\,861.33~\AA\ (in air) for the transition, with
Gaussian full width half maxima of 11.32~\AA\ and 11.147~\AA, respectively.}
\label{fig:hbeta_vel_profile}
\end{figure}

\begin{figure}
\includegraphics[angle=0,keepaspectratio,width=7.8cm]{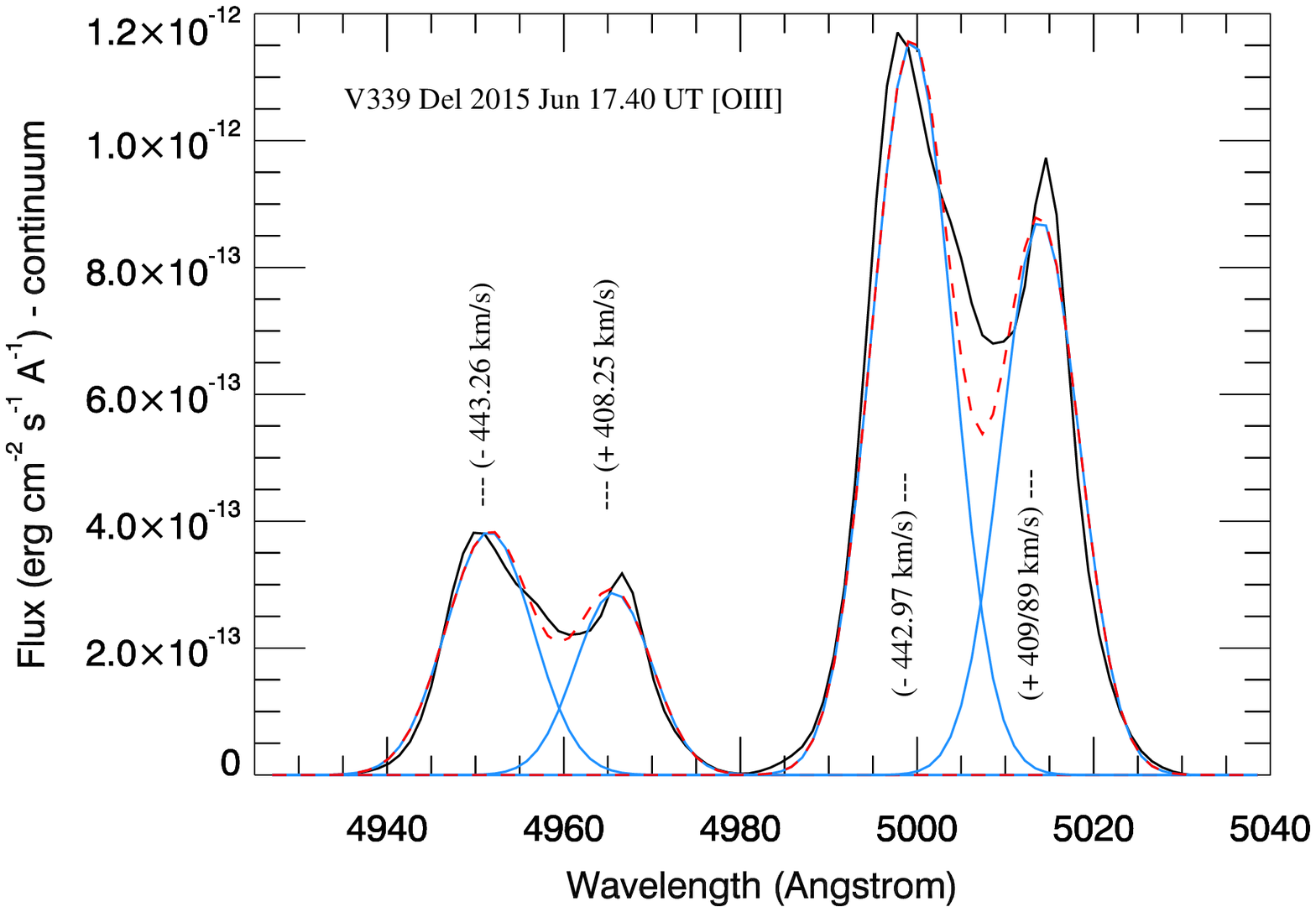}
\caption{The \fion{O}{iii} line profiles of V339 Del observed on 
2015 June 17.40 UT, extracted from the spectra shown in 
Fig.~\ref{fig:mmt_optical_spec}. The profile peaks exhibit the same structure 
as the Balmer series emission lines (Fig.~\ref{fig:hbeta_vel_profile}). 
Fitting a simple two-Gaussian model to deblend the velocity components of 
these forbidden lines returns components for the 4958.91 / 5006.84 lines
(wavelength in air) and Gaussian full width half maxima of
$-443.26$~km~s$^{-1}$, 11.55~\AA; $+408.25$~km~s$^{-1}$, 10.21~\AA\ and
$-442.97$~km~s$^{-1}$, 11.71~\AA; $+409.89$~km~s$^{-1}$, 10.35~\AA ,
respectively. The solid blue curves are the normalised Gaussian fits
to each component, the dashed red line is the composite sum of the normalised
Gaussian, and the solid black line is the original data.}
\label{fig:o3_vel_profile}
\end{figure}

\subsection{Optical Spectra}

By day 672 V339 Del had evolved into the nebular phase of its evolution. Our
optical spectroscopy (Fig.~\ref{fig:mmt_optical_spec}) show at this epoch emission
was dominated by recombination lines from the hydrogen Balmer series, helium ions and 
low ionization, forbidden line emission from metals such as C, N, O, Si, and Fe. The 
strongest emission lines in the spectra arise from \fion{O}{iii}4958.91/5006.84~\AA. 
The peak of the emission line profiles exhibit a ``double-horn'' structure 
indicating that the total observed emission at
a given wavelength arises from material receding from and approaching the observer (i.e.,
a ring or shell structure to the ejecta).

Using the IRAF SPLOT deblending tool, the 
components of line profiles were deblended assuming a simple model for the ejecta
geometry wherein two Gaussian components were fit to an emission line using
the average value of the local continuum using a 
non-linear least squares technique. 
Fig.~\ref{fig:hbeta_vel_profile} shows the structure of the optical Balmer H$\beta$ line
with the two fit velocity components, which are similar to those seen in the infrared
Paschen and Brackett series lines depicted in Fig.~\ref{VEL}(f) observed on day 683.06.
The spectral resolution of the optical and infrared spectra are comparable, and 
while the velocity peak of the blue shifted material is of the order $-500$~km~s$^{-1}$, 
that of the red shifted material differs by $\simeq 100$\vunit\ between the 
optical and the infrared profiles. However, this is less than the optical velocity 
resolution ($\approx 185$\vunit) hence we consider the two measurements comparable.

Emission from metal ions also dominates the optical spectra. The strongest of these lines
also exhibit emission from two velocity components as seen in the Balmer lines. 
Fig.~\ref{fig:o3_vel_profile} shows the Gaussian deblending of the 
\fion{O}{iii}4958/5006 \AA\ (rest wavelengths in air) forbidden
line profiles with the velocity components marked as well as each Gaussian fit. The
sum total of the two components are overlain on the observed emission profiles from the 
spectra for comparison.
Various lines were fit using this technique with integrated line fluxes determined from
measurements of the line cores and Gaussian full width half maxima. A detailed description
of line identifications derived from the optical spectra on day 672.50 -- uncorrected for
extinction -- is summarised in Table~\ref{line_ids_opt} in Appendix~\ref{APP}.

Detailed use of the optical and infrared line fluxes observed over multiple epochs
as input to photoionization abundance synthesis models is discussed in a forthcoming manuscript.

\section{Evolution of the dust \label{DUST}}

\begin{figure*}
\leavevmode
\includegraphics[angle=0,keepaspectratio,width=8cm]{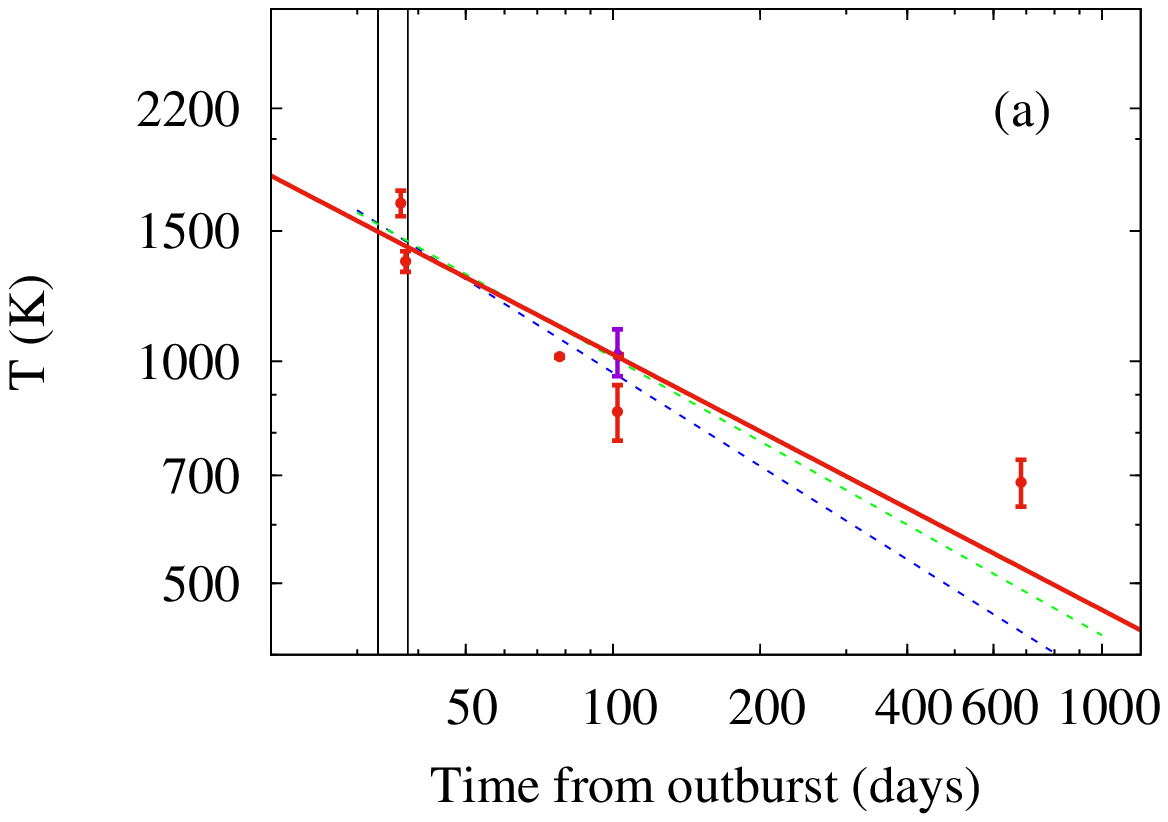}
\includegraphics[angle=0,keepaspectratio,width=8cm]{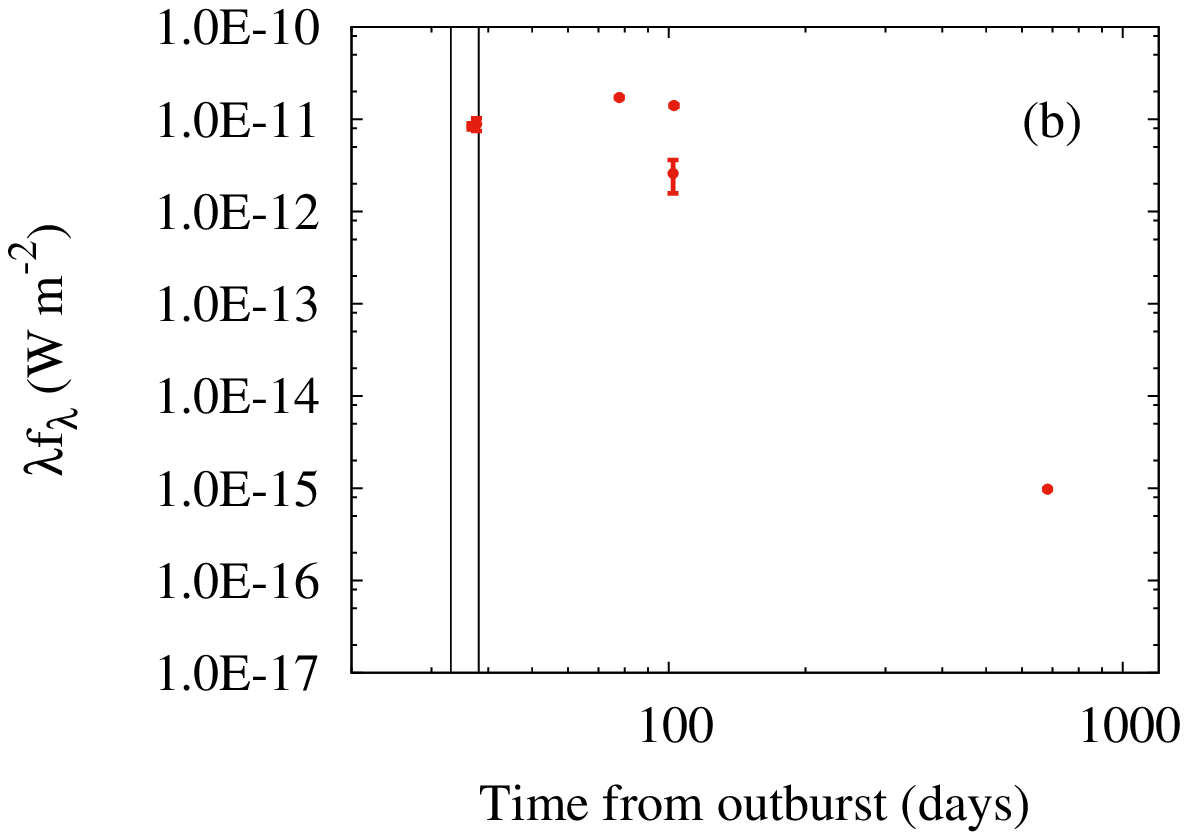}
\includegraphics[angle=0,keepaspectratio,width=8cm]{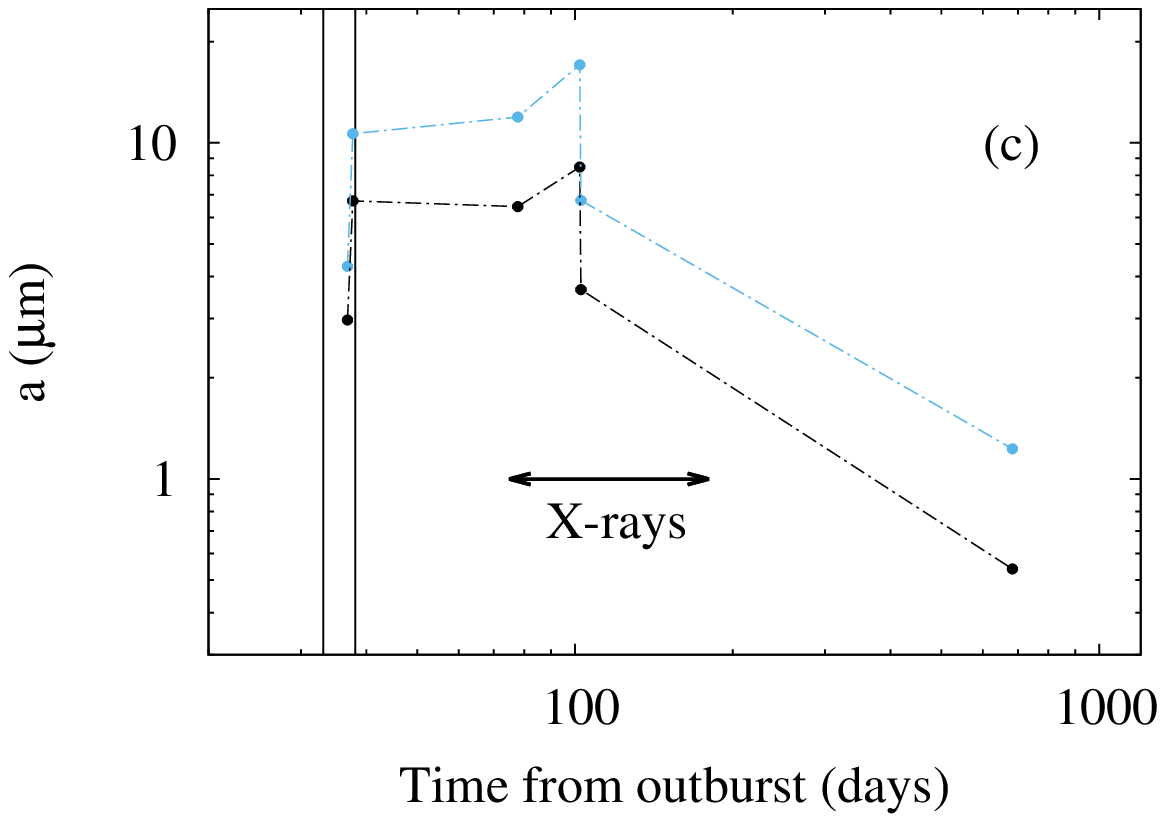}
\includegraphics[angle=0,keepaspectratio,width=8cm]{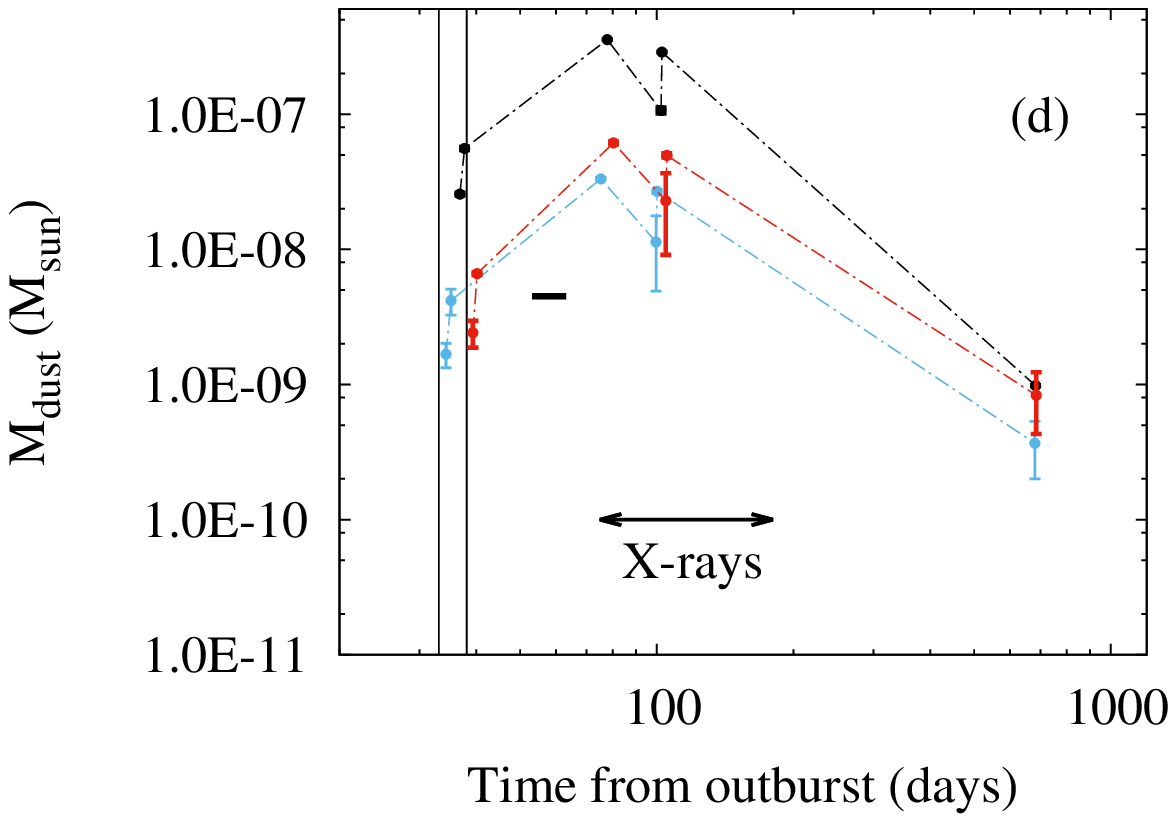}
\caption[]{Evolution of the dust; in each panel the vertical lines denote constraints on 
the epoch of dust formation. 
(a) Dust temperature; the magenta point at $t=102.14$~days is based on O'Brien Observatory
photometry. The solid line is a fit of $T\propto{t}^{-\sigma}$.
The broken green line is the expected relationship for graphitic carbon; 
the broken blue line is the expected relationship for amorphous carbon.
(b) Dust flux, given by $\{\lambda\,f_\lambda\}_{\rm max}$.
(c) Grain radius, for AC (black) and GR (blue) grains.
(d) Dust mass, calculated for 1\mic\ black body grains (black points), amorphous carbon
grains (blue points) and graphitic carbon grains (red points). The data for
amorphous carbon and graphitic carbon have been offset by --2.5~days and +2.5~days
respectively, for clarity, and dotted lines are included to guide the eye.
The short horizontal line depicts the dust mass reported by \cite{taranova}.
Uncertainties are smaller than the plotted points unless indicated. 
In panels (c) and (d) the approximate duration of the super-soft X-ray phase,
from Fig.~\ref{LC}(a), is indicated. See text for details.
\label{dust-evol}} 
\end{figure*}

The evolution of the IR spectral energy distribution is shown in Fig.~\ref{dust-fig}.  
On some of the epochs there is a clear
IR excess at the longer wavelengths, which we attribute to the 
dust reported by \cite{shenavrin} and \cite{gehrz-v339}; it is also evident that
there is a contribution from free-free and free-bound emission, although this 
becomes less important in the $K$-band. There is no evidence for the presence of dust
on or before 2013 September 16 (day 33.75; see Fig.~\ref{dust-fig}).

We attempted to fit a function of the form
\[ f_\lambda = \frac{C}{\lambda^{(5+\beta)}} \: \frac{1}{\exp[hc/\lambda{k}T_{\rm d}] - 1} \] 
to the excess. Here $C$ is a scaling factor, $\beta$ is the ``$\beta$-index'' for the dust,
defined in the usual way such that the dust emisivity is $\propto\lambda^{-\beta}$,
and $T_{\rm d}$ is the dust temperature. However, a non-zero value of $\beta$ persistently 
gave a poor fit for all datasets; best fits were obtained by
forcing $\beta\equiv0$, i.e. pure black body emission by the dust shell, although this does not
of course necessarily imply black body dust grains. The deduced black body temperatures
are typically in the range $\sim800-1600$~K. At these temperatures $f_\lambda$ peaks around
$1.8-2.8$\mic\ and the $K$-band data alone are sufficient to pin down the temperature;
this is particularly helpful as the free-free emission is (apart from the IRTF spectrum obtained
on day~683.06) less prominent at these
wavelengths. For example, fitting the black body function to both $H$- and $K$-band spectroscopic
data for day 102.66 gives a dust temperature of $1\,087\pm2$~K, compared with $1\,017\pm4$~K
for the $K$-band data alone (see Table~\ref{dust-tab}); the formal temperature
errors are small compared with $T$ as a consequence of the large number of data points in
the fit. In view of the uncertainties in the flux calibration
and in fitting the grain parameters (see below) we do not consider the difference 
to be significant.

We have therefore fitted the $K$-band data -- with emission lines removed -- with black bodies; 
the fits are shown in Fig.~\ref{dust-fig}. The evolution of the temperature
and dust flux (as given by $\{\lambda{f}_{\lambda}\}_{\rm max}$) are listed in Table~\ref{dust-tab} 
and shown in Fig.~\ref{dust-evol}(a) and (b) respectively. There is no clear evidence for the presence
of dust on 2013 September 16 (day 33.75; see Fig.~\ref{dust-fig}, top right panel); 
this is consistent with the conclusion
of \cite{taranova}, who found that dust formation occurred around 2013 September 17.
These authors determined a $(K-L)$ colour temperature of $\simeq1500$~K
on 2013 September 21, and $\simeq1200$~K on 2013 October 11, close to the values 
reported in Table~\ref{dust-tab}.

There remains a weak dust excess on day 683.06 but it is clear from Figs~\ref{IRTF1}
and \ref{dust-fig} that free-free and free-bound emission by the gas is comparable with,
if not dominating, the dust emission by this time. We have removed the free-free
and free-bound emission as determined in Section~\ref{IRTF2} to determine 
the dust excess, which has a black body temperature of 651~K.
There is no trace of dust emission shortward of 2.5\mic\ by day~839.30.

\subsection{The nature of the dust}

To determine the nature of the dust,
including dust mass and grain radius, we use the information in
Appendix~\ref{PLANCK}; this includes a revised Planck Mean absorption efficiency for
carbon dust,
determined from the data in \cite*{blanco}. The relevant information,
including formulae, is given in Appendix~\ref{PLANCK}.

\subsubsection{Grain radius} \label{grain_rad} 
To determine grain radius we use Equation~(6) of \cite{gehrz-v5668},
with $L_*=8.3\times10^5$\Lsun\ and $R=Vt$, where $V=583$\vunit\ (see above).
 
The dependence of grain radius on time is shown in Fig.~\ref{dust-evol}(c).
In reality of course, there will be a distribution of grain
sizes \citep[see e.g.][]{evans-cas} and the grain size calculated in this way
gives an average over the dust shell. However, it
seems clear that there is a rapid rise in the grain size immediately following grain
formation, eventually followed by a phase in which the grain size diminishes
substantially. This is further discussed below.

\subsubsection{Dust mass}

\paragraph*{Black body grains.}
For black body (BB) grains the dust mass is given by (see Equation~(\ref{BBMass}))
\begin{equation}
 \frac{M_{\rm d}}{\Msun} \simeq 2.19\times10^{15} \:\: \frac{\{\lambda{f}_{\lambda}\}_{\rm max}}{\mbox{W~m$^{-2}$}}
  \:\: \frac{1}{T_{\rm d}^{4}}
 \end{equation}
 where we have assumed 1\mic\ carbon grains ($\rho=2.25$~g~cm$^{-3}$) and distance $D=4.5$~kpc.
The dust masses are included in Table~\ref{dust-tab} in the column headed BB; the uncertainties 
have been propagated from the uncertainties in $\{\lambda{f}_{\lambda}\}_{\rm max}$ and $T_{\rm d}$. 
The dependence of dust mass on time is shown Fig.~\ref{dust-evol}(d).

\paragraph*{Optically thin amorphous carbon.}
For amorphous carbon (AC) grains (see Equation~(\ref{CCMass}))
\[ \frac{M_{\rm dust}}{\Msun} \simeq 3.76\times10^{17} \:\:  \frac{\{\lambda{f}_{\lambda}\}_{\rm max}}{T_{\rm d}^{4.754}} \]
where we have again taken $D=4.5$~kpc and grain density 2.25~g~cm$^{-3}$ for carbon.
The dust masses are again included in Table~\ref{dust-tab} (column AC) and
shown in Fig.~\ref{dust-evol}(d).

\paragraph*{Optically thin graphitic carbon.}
For graphitic carbon (GC) grains (again see Equation~(\ref{CCMass}))
\[ \frac{M_{\rm d}}{\Msun} \simeq 3.35\times10^{19} \:  \frac{\{\lambda{f}_{\lambda}\}_{\rm max}}{T_{\rm d}^{5.315}} \:\:\]
for $D=4.5$~kpc. The masses are given in Table~\ref{dust-tab} (column GR) and
shown in Fig.~\ref{dust-evol}(d).
As with grain radius, there seems to be evidence for a rapid rise in dust mass,
followed by a steep decline, irrespective of the composition of the dust.

\subsection{Grain temperature and mass}

Using the dust temperatures from Table~\ref{dust-tab}, we find that
the dust cools according to
\begin{equation}
 T_{\rm d} (\mbox{K}) = \frac{5032}{t^{0.346}(\mbox{d})} \:\:,
 \label{Tt}
\end{equation}
with an uncertainty of $\pm0.090$ in the exponent of $t$.
Grains having non-zero $\beta$, flowing at uniform velocity away from a heating source with
constant bolometric luminosity (as would be the case for a CN 
during the evolutionary phase
discussed here), would show a $T_{\rm d}\propto{t}^{-2/(\beta+4)}$ dependence;
the resultant $\beta=1.78^{+2.03}_{-1.19}$. This is slightly different from zero at the 
$\sim2\sigma$ level and hints at the presence of graphitic ($\beta\simeq1.32$) rather
than amorphous carbon grains 
($\beta\simeq0.75$; see also Fig.~\ref{dust-evol}(a)) but we consider various options below.

We recall that there was no evidence for dust on day~33.75 (see above), when the
dust temperature would have been $\simeq1\,490$~K according to Equation~(\ref{Tt});
\cite{taranova} concluded that dust formed on $\sim$~day 34.75, giving
$T_{\rm d}\simeq1\,470$~K. Extrapolation of Equation~(\ref{Tt}) leads us to conclude 
that the dust in \vd\ condensed at a temperature of $\simeq1\,480\pm20$~K.
Interestingly, this is well within the range of temperatures 
(1\,150 -- 1\,690~K) at which graphitic carbon condenses in carbon-rich flows 
\citep[][see also discussion in \cite{ER-CN}]{ebel}. This reinforces the 
conclusion that the dust in \vd\ was graphitic carbon.
Furthermore, it is of interest to note that the dust in the classical nova V2326~Cyg
condensed at a temperature $1\,410\pm15$~K \citep{lynch2}, close to the value
of the dust condensation temperature we have deduced for \vd.

There is further, albeit very circumstantial, evidence for carbon dust in the broadband 
photometry from the O'Brien Observatory, obtained on day 102.14.
In these data there seems to be an excess with respect to the
black body continuum at $\sim11-12$\mic\ (see Fig.~\ref{dust-fig}),
although there was no evidence for a feature in this wavelength range in the SOFIA 
data on day~27.4 \citep{gehrz-v339}.
There are a number of possible sources for this excess
on day~102.14. The most plausible assignment would be Hu$\alpha$ at 12.3719\mic\
(\fion{Ne}{ii} 12.8135\mic, although sometimes present even in CO novae \citep[see e.g.][]{evans-2},
would need to be implausibly strong to have a significant effect on broadband data).
Another alternative, which would be consistent with the presence
of carbon dust, is the 11.1/12.7\mic\ C\chemone{H} out-of-plane bending modes in 
polycyclic aromatic hydrocarbons, commonly seen in nova dust 
\citep[see e.g.][and references therein]{helton-eas}.

Irrespective of dust composition, there seems to be some evidence for an increase 
in dust mass up to $\sim$~day 100, followed by a substantial decline to day $\sim680$. 
The increase may be attributed to grain growth, an increase
in the number of emitting grains, or a combination of both these effects. 
Conversely the subsequent decline must 
be due to a decline in grain size or decrease in grain number, either of which points to 
grain destruction on a substantial scale.

\subsection{A dust minimum in the visual light curve}

It is of interest to correlate the evolution of the dust with the behaviour of
the visual light curve.
Following visual maximum, the light curve declines uniformly, but the decline becomes
distinctly slower after $\sim{\mbox{JD}}2\,456\,530$ ($\sim$~day~12;
see Fig.~\ref{LC}(a)); then there is a distinct
dip in the light curve as dust forms. We have fitted the light curve immediately before 
and after the dust dip with a function of the form
\[ m_{\rm v} = A + B\:(\mbox{JD}-2455000) \]
and find $A=-95.94\pm2.09$ and $B=0.0669\pm0.0013$~day$^{-1}$ (note that $B$ must not
be associated with the ``speed class'' of the nova). We rectified the light curve
with this decline to highlight the dust dip (see Fig.~\ref{LC}(b), in which the dust optical
depth $\tau=\Delta{m}_{\rm vis}\times0.4\ln{10}$ is plotted against the Julian Date).
There is a clear maximum in $\tau$, with $\tau\simeq0.7$, around JD 2\,456\,580
($\sim$~day~62).
Within the limitations of the cadence of the IR observations, this coincides with maximum grain
size and maximum grain mass in Fig.~\ref{dust-evol}(c,d); the duration of the dust phase, as
determined by the IR observations, is consistent with the width of the dust dip in Fig.~\ref{LC}(b).

\subsection{Grain destruction}
\label{dest}
Evidence for grain destruction in a nova wind was first noted in the case of nova LW~Ser (1978)
by \cite{gehrz-ser}. The physical processes associated with grain 
destruction in nova winds have been discussed by \cite*{MEB}, \cite{ME}, \cite*{MEA} and \cite{rawlings}, in
the context of sputtering and chemi-sputtering of carbon dust,
and the effect of annealing carbon dust in the
UV radiation field of the nova. 

However, more recently the effect of X-radiation
on the survival of dust has been the subject of considerable interest 
in the context of $\gamma$-ray bursts \citep{fruchter}, and
as \vd\ was an X-ray source (see Section~\ref{intro}), it is of interest to consider the
effect of hard radiation on the survival potential of dust in the environment of \vd.

\cite{fruchter} considered the competition between the input of energy into grains
by X-radiation, and energy loss by grain radiation and sublimation.
They also considered the effect of grain shattering due to electrostatic stress,
resulting from grain charging following the ejection of electrons from the $K-$shell of
atoms within the grain.

We see from Fig.~\ref{dust-evol} that, when the dust mass peaked  around day 100, the dust
temperature was $\sim1\,000$~K and steadily declined thereafter; is seems unlikely 
therefore that, at $T\ltsimeq1\,000$K, graphitic grains -- for which the sublimation 
temperature is $\gtsimeq1\,800$~K for carbon-rich environments \citep{lodders} -- 
would be subject to evaporation: it is more likely that it is grain shattering by 
electrostatic stress that destroyed the grains around \vd.

\cite{fruchter} found that, whether or not a grain survives following exposure
to X-radiation is determined by the
parameter $E_{51}/D_{100}^2$, where $E_{51}$ is the energy radiated in the form of
X-rays at 1~keV, normalised to $10^{51}$~ergs, and $D_{100}$ is the distance of the grains
from the X-ray source in units of 100~pc;
from Figure~1 of \citeauthor{fruchter}, grains are shattered if
$E_{51}/D_{100}^2\gtsimeq1$.

We suppose that, at the time of maximum
grain size, the nova was a SSS with the bulk of the radiation being emitted at
X-ray wavelengths. Assuming that the grains are exposed to X-rays for a time
$\Delta{t}$, at time $t$ after the eruption, the \citeauthor{fruchter} parameter is
\[ \frac{E_{51}}{D_{100}^2} \simeq 1.7\times10^3 \: \Phi \: \left ( \frac{L_*}{\Lsun} \right) \:
    \left( \frac{\Delta{t}}{\mbox{days}} \right) \: \left(\frac{t}{\mbox{days}}\right)^{-2} \: \left(\frac{V}{500\vunit}\right)^{-2} \]
where $\Phi$ is the fraction of the nova radiation emitted at 1~keV.
Taking $L_*\simeq8.3\times10^5$\Lsun\ \citep{gehrz-v339}, $V=583$\vunit, $t\simeq100$~days,
$\Delta{t}\simeq100$~days, we find $E_{51}/D_{100}^2\sim10^6\Phi$.
So if $\Phi\gtsimeq10^{-5}$, which seems reasonable,
charging of grains by X-radiation 
is more than sufficient to shatter the grains.

However, while it seems that exposure to X-radiation would
destroy the dust, we should express a note of caution in that the 
\citeauthor{fruchter} analysis assumes
a power-law for the X-ray source; while this would be valid for the
hard X-radiation, it is less so for the near-black body X-ray
source appropriate for a CN during the SSS phase (even though
pure black body emission does not generally provide a good description
of the SSS phase of novae). 
This issue will be addressed in a forthcoming paper.

Finally, we speculate that the shattering of $\sim$ micron-sized graphitic grains
in the way we have suggested might 
result in the release of significant amounts of polycyclic aromatic hydrocarbon 
(PAH) fragments, possibly even fullerenes, into
the nova environment. If this is the case classical novae may display persistent 
PAH emission \citep[see][for a summary of PAH emission in novae]{helton-eas} for some 
time after the dust formation phase has apparently come to an end.

\section{Conclusions}

We have presented infrared spectroscopy of the classical nova \vd.

The IR spectrum is initially dominated by emission by the ejected gas,
with \pion{H}{i} and low excitation atomic CNO lines being prominent.
There is clear evolution in the emission line profiles. The lines
are initially ($t\ltsimeq37$~days) symmetric, with HWHM $\sim530$\vunit.
However, after day~77, they become highly asymmetric, with a strong blue
and weak red wing. Later still ($t\gtsimeq600$~days) the emission lines
display a castellated structure.

Rapid dust formation occurs around day~34.75, following which 
the IR emission becomes dominated by the dust.
The dust condensation temperature was $1\,480$~K, consistent with
the notion that the dust is graphitic. We find that $\sim5\times10^{-9}$\Msun\ of
dust was formed, and that the grains grew to a dimension of a few \mic.
We further find that both
the mass of dust, and the radius of the dust grains, increased rapidly
following the formation of dust, peaked around 100~days after eruption,
and thereafter declined precipitously. We attribute this to the charging
of dust grains by the X-ray emission of \vd, causing
the grains to shatter due to electrostatic stress.

\section*{Acknowledgements}

We thank Dr Kim Page for valuable information about the early X-ray emission of
\vd, and the British Astronomical Association and The Astronomer magazine
for providing its visual light curve. 
We also appreciate the referee's supportive comments on this paper.

RDG acknowledges support from the National Aeronautics and Space
Administration (NASA) and the United States Air Force.
VARMR acknowledges financial support from the Radboud Excellence Initiative.
CEW acknowledges partial support from NASA (HST-GO-13828.008-A).
SS gratefully acknowledges partial support from both NASA and NSF
grants to ASU.
The research work at the Physical Research Laboratory is supported by the 
Department of Space, Government of India. DS is a visiting astronomer at 
the Infrared Telescope Facility which is operated by the University of 
Hawaii under contract NNH14CK55B with NASA.





\include{v339_bib_v1}


\appendix

\section{Observed line fluxes \label{APP}}

\subsection{Mt Abu data}
The line centres and fluxes were determined by subtracting a continuum
and fitting Gaussian functions to the emission line profiles.
The uncertainties in the line centres and fluxes arise from the placement of the continuum
and the Gaussian fitting. Repeated experiments with slight variations in the assumed
continuum, and comparison with line fitting on different dates, showed that the 
uncertainties in the line centres are typically
$\pm0.0004  -0.001$\mic, and $\pm5$\%  to $\pm12\%$ in the line fluxes.

The line fluxes are listed in Table~\ref{line_ids_e}, which lists data to day 26.70,
and Table~\ref{line_ids_l}, which lists data from day 27.76 onwards.
Both Tables give measured wavelengths, suggested
identifications and transitions.

\input{line_ids_early}
\input{line_ids_late_rv}

\subsection{MMT data}
Line fluxes and centres for each component associated with a given line identification 
are summarised
in Table~\ref{line_ids_opt}. The components of line profiles where deblended assuming a simple model for the ejecta
geometry wherein two Gaussian components where fitted to an emission line using a
non-linear least squares technique and the average value of the local continuum. For the Balmer
H$\alpha$ line region, multiple Gaussians were used in an attempt to deblend and fit a rather complex 
emission profile. Repeated fitting, with the continuum and gaussian full width half maximum as
free parameters, suggests that the line profile centers are accurate to $\approx \pm 0.005$\AA,
with fluxes $\ltsimeq \pm 5$\% for most lines, except H$\alpha$ wherein the uncertainies are $\approx 10$\%.
Tentative line identifications and transitions cited in the table are guided by expected emission 
lines commonly seen in novae in the nebular stage of evolution 
\citep[e.g.,][]{2012AJ....144...98W}, ionisation states (eV), and atomic line lists 
contained photoinoization codes\footnote{see: http://www.pa.uky.edu/\~{}peter/newpage/}.

\input{v339del_A1_optical_line_fluxes1}

\section{Planck Means and dust properties} \label{PLANCK}

\begin{figure*}
\begin{center}
\leavevmode
\includegraphics[angle=0,keepaspectratio,width=7cm]{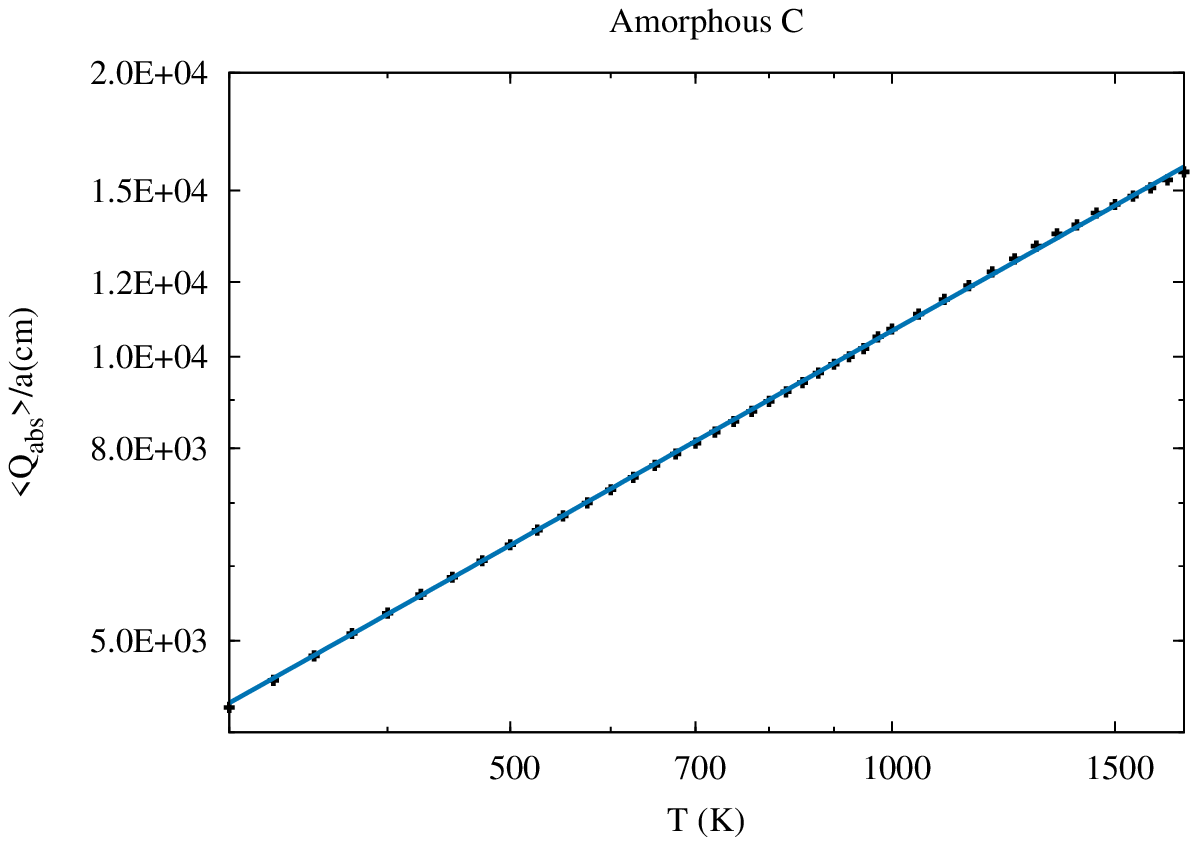}
\includegraphics[angle=0,keepaspectratio,width=7cm]{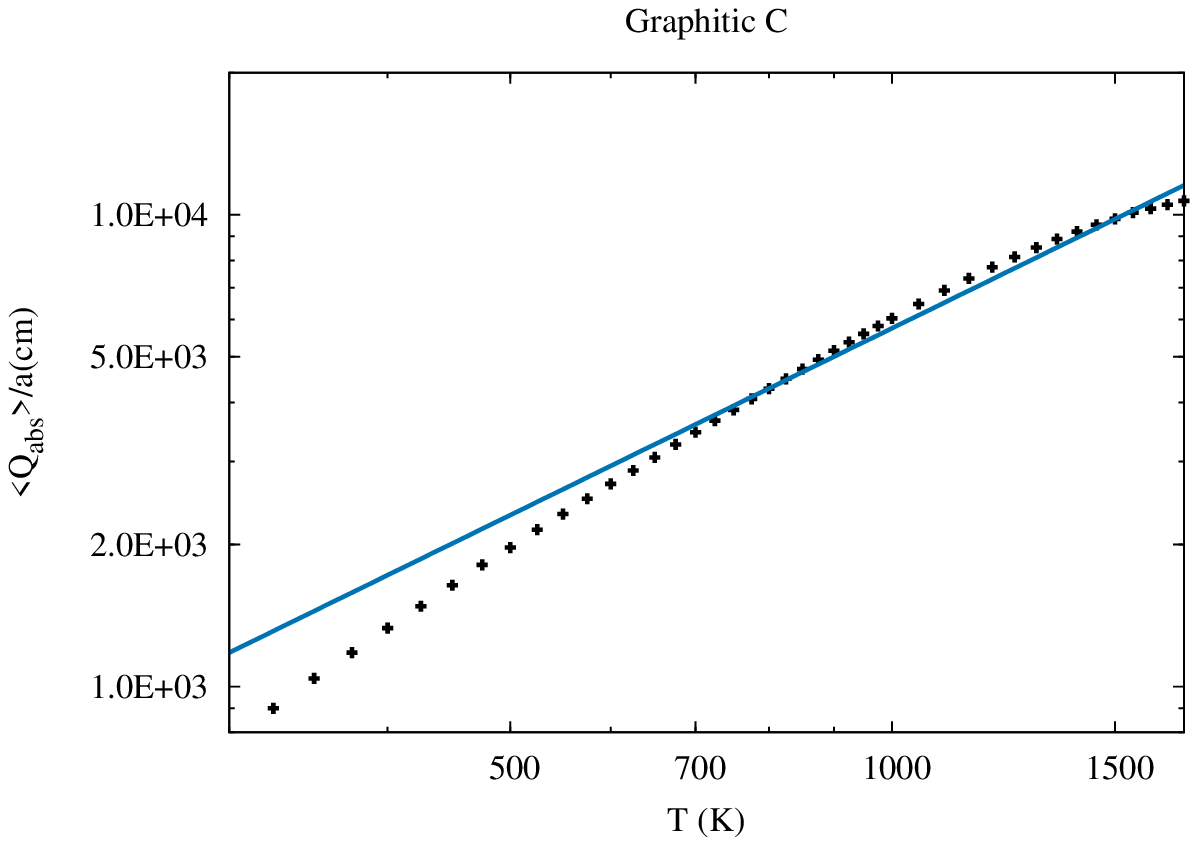}
\caption[]{Fit of functions of the form $\langle{Q}_{\rm abs}\rangle=AaT^\beta$ to
laboratory Planck mean data in \cite{blanco}. Left, amorphous carbon,
right, graphitic carbon. The fits are confined to the 
temperature range 400 -- 1700~K. \label{FITS}} 
\end{center}
\end{figure*}

In this section we summarise our determination of the Planck mean absorption
efficiencies of carbon dust, together with the determination of dust masses
for the case of dust shells that are optically thin in the IR.

\subsection{Planck means}

Most previous estimates of dust masses in CNe \citep[see e.g.][and references therein]{gehrz-CN,EG-BASI}
are based on the Planck means plotted
by \cite{gilman}. These in turn are based on a variety of old data and the Planck
means are in need of updating. We have used the Planck means measured and 
tabulated by \cite{blanco}.  These have been compiled in a systematic
fashion and cover the temperature range of interest for CNe.

For carbon, the dependence of the Planck mean absorption efficiency $\langle{Q}_{\rm abs}\rangle$
on temperature $T$ follows, to a good approximation, a power law of the form
\[ \langle{Q}_{\rm abs}\rangle \simeq Aa^\gamma{T}^\beta \:\:,\]
where $A$ is a constant, $a$ is the radius of the (spherical) grain and $\beta$ is
the $\beta$-index for the dust. For carbon grains $\gamma\equiv1$.

For amorphous carbon (AC) grains, we take Planck mean absorption efficiencies 
from \cite{blanco}, using their ``TU'' sample. We find that 
\[ \langle{Q}_{\rm abs}\rangle =  58.160 \: a \: T^{0.754} \:, \]
with $a$ in cm, $T$ in K, provides an excellent fit over the temperature range 400--1\,700~K; 
$A=58.160\pm0.754$ and the $\beta$-index is $0.754\pm0.003$.

For graphitic carbon (GC) we again take the Planck mean absorption efficiencies from \cite{blanco};
these are reasonably fitted over the range 400--1\,700~K by
\[ \langle{Q}_{\rm abs}\rangle = 0.653 \: a \: T^{1.315} \:, \]
with $a$ again in cm and $T$ in K; $A=0.653\pm0.128$ and the $\beta$-index is $1.315\pm0.027$.

Both fits are shown in Fig.~\ref{FITS}. The AC is clearly well described by a power law
but the power law fit to the GR data is rather less stisfactory, particularly at 
lower temperatures; this is evidenced by the uncertainties in the respective values of
$A$ and $\beta$. The temperature-dependence for GR is steeper at lower temperatures,
and levels off at higher temperatures. Indeed this tendency is clearly evident in
\citeauthor{gilman}'s (1974) Figure~4.
However, the assumption of a power law dependence leads to a straight-forward and
easily-applied expression for the dust mass and grain radius and, as the deduced dust 
temperature is at the higher end of the range, we use this simple form here.

\subsection{Dust mass}

\subsubsection{Carbon grains}

For carbon grains the dust emission in the IR, integrated over wavelength, is given by
\begin{equation}
 f  =  \frac{{a}^2}{D^2} \: N_{\rm gr} \: \frac{\sigma{T}^4}{\pi} \: \langle{Q}_{\rm abs}\rangle
   = \frac{{a}^2}{D^2} \: \frac{3M_{\rm d}}{4\pi{a}^3\rho} \:\: \frac{\sigma{T}^4}{\pi} \:\:  AaT^\beta \:\:, \label{flux-c}
\end{equation}
when the dust shell is optically thin in the IR; the number of emitting grains is
\[ N_{\rm gr} = \frac{3M_{\rm d}}{4\pi{a}^3\rho} \:\: \]
and $\rho$ is the density of the grain material. The observed parameter
is $\{\lambda{f}_\lambda\}_{\rm max}$ which is related to $f$ by
\[ f = 1.359\ldots \{\lambda{f}_\lambda\}_{\rm max} \:\:.\]
\citep[see][]{G-N}. Rearranging, and using $\rho=2.25$~g~cm$^{-3}$ for carbon, gives
\begin{eqnarray}
 M_{\rm d} & = & 1.359 \: \{\lambda{f}_\lambda\}_{\rm max} \: \frac{D^2}{3} \: \frac{4\pi\rho}{A\sigma{T}^{4+\beta}} \nonumber \\
 \mbox{and~~~~~} \frac{M_{\rm d}}{\Msun}   & = &  
     1.08\times10^{20} \: \left ( \frac{\{\lambda{f}_\lambda\}_{\rm max}}{\mbox{W~m$^{-2}$}} \right ) 
                               \left ( \frac{D}{\mbox{kpc}} \right )^2 \:\: \frac{1}{AT^{(\beta+4)}} \:\:, \nonumber \\
                        \label{CCMass}       &&
\end{eqnarray}
independent of $a$.

\subsubsection{Black body grains}

In this case $A=1$, $\beta=\gamma=0$ and the equivalent of Equation~(\ref{flux-c}) for
the observed dust emission is given by
\begin{equation}
 f = \frac{\pi{a}^2}{D^2} \: N_{\rm gr} \: \frac{\sigma{T}^4}{\pi} \:\:, \label{flux}
\end{equation}
Rearranging gives
\begin{eqnarray}
 {M_{\rm d}}  & = & 1.812 \: D^2  \: \{\lambda{f}_\lambda\}_{\rm max} \:\: \frac{\pi{a}\rho}{\sigma{T}^4}  \: \nonumber\\
 &&\nonumber\\
    \mbox{and~~~~} \frac{M_{\rm d}}{\Msun} & \simeq & 4.805\times10^{16} \:\: 
           \frac{a \rho \, D_{\rm kpc}^2 \, \{\lambda{f}_{\lambda}\}_{\rm max}}{T^{4}}\:\:. \label{BBMass}  
\end{eqnarray}
Note that in this case $M_{\rm d}$ depends on $a$.

\bsp	
\label{lastpage}
\end{document}

%% file: line_ids_early.tex
\setcounter{table}{0}
\begin{table*}
 \centering
\begin{minipage}{180mm}
  \caption{Line fluxes measured from Mt Abu data. Earlier data, days 4.73 to 26.70. Fluxes are given in units of $10^{-14}$~W~m$^{-2}$ and
  are uncorrected for extinction.
  Uncertainties in the last one, two or three digits are given in brackets; thus 13.0(15) means $13.0\pm1.5$,
  469(11) means $469\pm11$ etc. \label{line_ids_e}}
   \begin{tabular}{cccccccccccc}
  \hline
  $\lambda_{\rm obs}$ & ID, $\lambda_0$ & Transition       & \multicolumn{9}{c}{Line flux ($10^{-14}$~W~m$^{-2}$) on days given}    \\ \cline{4-12}
       ($\mu$m)       &     ($\mu$m)    &   $\ell-u$       & & & & & & & & &  \\ 
      &                    &                       &$t=4.73$&14.76&15.79&19.78&20.73&22.73&24.72&25.75&26.70\\ \hline
1.0828&\pion{He}{i}, 1.0833&$^3$S$-^3$P$^o$        & -- & --    & --  & --    &  --     & --    &  --      & --    & -- \\
1.0936&\pion{H}{i},  1.0941&3--6                   & -- & 153(5)& --  & --    &82.7(36)&109(5)&99.3(57)&105(7)& -- \\
1.0946&\pion{N}{i},  1.0946&$^4$S$^o_{3/2}-^4$P    & -- & --    & --  &       & --      & --    & --       & --    &122(95)\\
1.1284&\pion{C}{i},  1.1281&$^1$D$_2-^1$F$^o_3$    & -- &  --   & --  &  --   & --      & --    &  --      & 470(8)& --\\
1.1293&\pion{O}{i},  1.1290&$^3$P$-^3$D$^o$        & -- & --    & --  & 320(5)&313(4)   &456(6) &392(7)    & --    & 469(11) \\
1.1300&\pion{N}{i},  1.1297&$^4$D$^o_{1/2}-^4$P$_{1/2}$
                                                   &--  &353(8) &433(9)& --   & --      & --    & --       & --    & -- \\
      &\pion{O}{i},  1.1301&$^5$P$_2-^5$S$^o_2$    & -- & --    & --   & --   & --      & --    & --       & --    & -- \\
      &\pion{C}{i},  1.1301&$^3$P$^o_1-^3$D$_2$    &  --& --    & --   & --   & --      & --    & --       & --    & -- \\
      &\pion{N}{i},  1.1301&$^4$P$^o_{1/2}-^4$P$_{1/2}$
                                                   & -- & --    & --   & --   & --      & --    & --       & --    & --  \\
      &\pion{N}{i},  1.1300&$^4$P$^o_{5/2}-^4$D    & -- & --    & --   & --   & --      & --    & --       & --    & --  \\
      &\pion{N}{i},  1.1303&$^2$P$^o_{1/2}-^2$D    & -- & --    & --   & --   & --      & --    & --       & --    & -- \\
      &\pion{N}{i},  1.1304&$^4$D$^o_{3/2}-^4$D$_{1/2}$
                                                   & --  & --   & --   & --   & --      & --    & --       & --    & --  \\
1.1463&                    &                       & --  & --   & --   & --   & --      & --    & --       & --     & -- \\
1.1619&\pion{C}{i}, 1.1618&$^3$D$-^3$D$^o$         & --  & --   &24.2(12)&9.17(69)& --  & --    &3.92(50) & --     & --  \\
      &\pion{C}{i}, 1.1622&$^3$D$-^3$D$^o$         & --  & --   & --     & --     &  --    & --     & --       & --      & -- \\
1.1654&\pion{C}{i}, 1.1651&$^3$D$_2-^3$D$^o_1$     & --  &75.2()& --     &29.3(36)&27.9(16)&35.0(92)&24.2(1.63)&26.4(1.3)&31.7(94)\\
1.1662&\pion{C}{i}, 1.1662&$^3$S$-^3$P$^o$         &--   &      &48.0(58)& --     & --     & --     & --       & --     & -- \\
1.1758&\pion{C}{i}, 1.1758&$^3$D$_2-^3$F$^o_3$     & --  &151(3)&160(4)  &91.5(18)&76.5(35)&92.5(18)&73.7(11)  &74.9(14)&80.4(10) \\
      &\pion{C}{i}, 1.1757&$^3$D$_3-^3$F$^o_4$     & --  & --      & --  & --    & --      & --     & --      & --       & -- \\
1.1863&\pion{C}{i}, 1.1864&$^3$P$_2-^3$D$^o_3$     & --  &49.4(25) & --  & --     & --     & --     & --      & --  & --\\
      &\pion{C}{i}, 1.1864&$^3$P$_2-^3$D$^o_2$     & --  & --      & --  & --    & --      & --     & --      & --  & --\\
      &\pion{C}{i}, 1.1866&D$_1-^3$P$^o_1$         &--   & --      & --  & --    & --      & --     & --      & --   & --\\
      &\pion{N}{i}, 1.1861&$^4$S$^o_{3/2}-^4$P$_{3/2}$ & --&  --   & --  & --    & --      & --     & --      & --   & --\\
1.1887&                   &                        & --    & --  & --    & 16.3(13)&18.2(18)&21.9(11)&16.8(7)&13.7(9)&19.7(10) \\
1.2087&                   &                        &  --   & --  & --    & --      & --     & --     & --    &   --  &12.5(14)\\
1.2191&                   &                         & --  & --    & --     & --      & --     & --    & --    & -- &4.19(34) \\
1.2300&                   &                        & --   & --    & --    & --     &    --   &   --  &   --   & --  & 8.76(51)\\
1.2466&\pion{N}{i}, 1.2473&$^2$D$^o_{5/2}-^2$F$_{7/2}$&-- &26.4(52)&31.8 (37)&10.5(13)&11.4(23)&11.7(8)&11.5(15)&11.0(14)&11.2(10) \\
\hline
  \end{tabular}
  \end{minipage}
\end{table*}

\setcounter{table}{0}
\begin{table*}
 \centering
\begin{minipage}{180mm}
  \caption{Continued. Line fluxes measured from Mt Abu data. Earlier data, days 4.73 to 26.70. Fluxes are given in units of $10^{-14}$~W~m$^{-2}$
  and   are uncorrected for extinction.
  Uncertainties in the last one, two or three digits are given in brackets; thus 13.0(15) means $13.0\pm1.5$,
  469(11) means $469\pm11$ etc.}
   \begin{tabular}{cccccccccccc}
     \hline
  $\lambda_{\rm obs}$ & ID, $\lambda_0$ & Transition       & \multicolumn{9}{c}{Line flux ($10^{-14}$~W~m$^{-2}$) on days given}    \\ \cline{4-12}
       ($\mu$m)       &     ($\mu$m)    &   $\ell-u$       & & & & & & & & &  \\ 
      &                    &                       &$t=4.73$&14.76&15.79&19.78&20.73&22.73&24.72&25.75&26.70\\ \hline
1.2509&                   &                        & --   & --     & --      & --     & --     & --    & --     & --     & -- \\
1.2586&\pion{N}{i}, 1.2586&$^2$F$_{7/2}-$D$^o$     & --   &46.5(10)&49.1(11)&17.6(16)&17.3(28)&17.6(10)&17.1(38)&16.5(4)&14.7(10) \\
      &\pion{C}{i}, 1.2585&$^3$P$_1-^3$P$^o_2$     & --   & --     & --     & --     & --      & --    & --     & --    & --\\
      &\pion{N}{i}, 1.2585&$^2$D$^o_{5/2}-^2$F$_{5/2}$& -- & --    & --     & --     & --      & --    & --     & --     & --\\
1.2809&                   &                        & -- & -- & --& --& --& --& --& --& --\\
1.2823&\pion{H}{i}, 1.2822&3--5                   &8.40(107)&213(3)&261(4)   &175(2)&170(3)&198(2)&202(2)&218(2)&224(2) \\
1.2975&\pion{N}{i}, 1.2973&$^2$P$-^2$D$^o$         & --    &       &3.582(32)&   --   &  --   & --    & --      & --   & -- \\
1.3172&\pion{N}{i}, 1.3177&$^2$D$_{3/2}-^2$D$^o_{5/2}$&--&29.0(6)&   --    & 14.3(14)&11.9(23)& --  & 13.0(15)&13.0(15)&15.2(18) \\
      &\pion{N}{i}, 1.3176&$^2$D$_{5/2}-^2$D$^o_{5/2}$ & --& --    & --      & --      & --     & --  & --      & -- & -- \\
   &&&&&&&&&&&\\
1.5268&                   &                          & --    & --      & --     &1.29(9)  & --   & --     & -- & -- &1.77(10)\\
1.5348&                   &                          & --    & --      & 3.46(20)&1.82(11)& --   & --     &1.81(8)& --&1.77(9) \\
1.5447&                   &                          & --   & --       &4.84(20)&2.80(12)& --    & 3.06(14)&2.56(8)&2.54(14)&2.33(9) \\
1.5568&\pion{H}{i},1.5561 & 4--16                    &--    &9.23(28)  &9.64(22)&5.56(13)& --    &6.33(15)&5.92(10)&6.18(17)&5.67(11)\\
1.5707&\pion{H}{i},1.5705 & 4--15                    &--    &7.88(31)  &9.07(24)&5.14(14)& --    & 5.82(16)&5.39(10)&4.99(17)&4.58(10)\\
1.5748                    &                         & --    & --       & --     & --      & --   & -- &  --  &  --  & --   & --\\
1.5789&                   &                         & --    & --       & 1.47(16)&0.71(9)& --    &0.69(7) &0.58(7)&0.55 (7)& -- \\
1.5847&\pion{H}{i},1.5885&4--14                   &--     &11.9(3)  &12.3(2)  &6.63(13)& --    &6.55(15)&6.19(9)&5.95(16)& -- \\
1.6021&                   &                        & --   &7.68(28)&7.75(23)&3.37(9)& -- &3.12(7)&2.89(7)&2.85(7)&2.95(16)\\
1.6054                    &                         & --    & --       & --     & --      & --   & -- &  --  &  --  & --   & --\\
1.6115&\pion{H}{i}, 1.6114& 4--13                 &--     &10.7(2)&11.3(2)  &6.58(14)& -- &6.63(15)&6.87(10)&7.05(17)&6.69(17)\\
1.6329&                   &                       &--     & --    & --       &0.44(9)& -- & --     & --     & --     & --\\
1.6142                    &                       & -- & -- & --& -- & -- & --&  --& -- &-- & --\\
1.6416&\pion{H}{i}, 1.6412&4--12                 &--      &21.8(5)&21.67(113)&10.2(3)& -- &10.5(6) &10.4(4 )& 10.3(2)&10.5(2)\\
1.6499&                    &                     & --     & --    & --       & --    & --  &0.69(7)& --     & -- & --\\
1.6711&                    &                     & --    &  --    & --       &0.95(7)&  -- & --    & --     & --  & --\\
1.6808&\pion{H}{i}, 1.6811&4--11                 & --    &19.7(31)&15.3(7)   &9.38(28)&-- &10.5(4)& 10.6(3)&10.2(3)&11.7(6)\\
1.6826&                   &                      & --     & --      & --  & --         & -- &     --&  -- & -- & --\\
      1.6892&\pion{C}{i}, 1.6895&$^1$D$_2-^1$F$^o_3$   &2.69(25)&28.0(29)&44.4(10)& --    & --   &22.1(9)&23.3(4)&21.9(5)&18.8(5)\\
1.7058&                   &                      &--      & --     &  --    &1.57(9)& --   & --    & --     & --&  --\\
1.7244&                    &                    & --     & --      & --     & --    & --   & --    & --     & --   & --\\ 
1.7251&                   &                       & --   & --     & --      &3.71(25)& --  &3.59(39)&3.29(41)&2.71(30)&2.72(16) \\
1.7362&\pion{H}{i}, 1.7367 &4--10                 &--    & --     & --      &22.5(4) &  -- &25.4(11)&24.7(5)&23.1(6)&19.0(6)\\
1.7450&                      &                     & --   & --     & --      & --     & --   &13.0(5)& --     &11.0(2)& -- \\
1.7563&                   &                       & --   & --     & --      & --     & --   &4.53(29)& --    &4.75(21)& --\\
1.7667&                   &                       & --   & --     & --      & --     & --   &11.6(10)&9.52(50)&7.98(63)&--\\
&&&&&&&&&&&\\
1.9741&                     &                       &--    & --     & --      &1.71(6) & --  &1.25(6)&2.19(10)&1.43(4)& --\\
2.0599&\pion{He}{i}, 2.0587&$^1$S$_0-^1$P$^o_1$ &   --    & --    & --       &1.546)&2.13(9)&1.26(9)&1.77(8)&1.90(8)&1.87(7) \\
2.0757 &                     &                  &--        & --    & --       &0.32(4)   &0.37(7)& --   & --    & --  & --\\
2.0890&                      &                   & --      & --     & --       &0.47(0.04)&0.70(8)&--     &0.66(4)&0.54(4)&0.81(4) \\
2.1029&\pion{C}{i}, 2.1029&  $^1$S$-^1$P$^o$    &--       &1.50(14)&2.21(17)  &0.66(4)   &0.74(5)& --    &0.45(3)&0.35(3)&0.43(4)\\
2.1235&                   &                     &--       &5.85(23)&7.34(13)  &2.81(6)   &3.07(7)&2.56(9)&2.76(7)&2.24(5)&2.13(6)\\
2.1492&                   &                     &--       & --     &3.32(12)  & --       & --     &1.28(7)&1.33(6)& -- &0.76(3)\\
2.1505&                   &                     & =[ii0--      & --     & --       & --       &1.46(6) & --     & --   &1.01(4)& --\\
2.1645                    &                         & --    & --   & --     & --      & --   & -- &  --  &  --  & --   & --\\
2.1667&\pion{H}{i}, 2.1661&4--7                   &--- &22.5(35) &20.9(2)& -- &15.3(2)&14.1(2)&18.0(2)&15.5(2)&16.6(3) \\
2.2153&                   &                       &--    &2.44(18)&1.87(17)&0.56(5) & -- & -- &0.41(3)&0.28(3)&0.44(5)\\
2.2891&                    &                     &--     &1.17(13)& --      & --    & --  & --  & --   & -- & --\\
2.2918&\pion{C}{i}\, 2.2913&$^1$S$_0-^1$P$^o_1$   &--      & --     &2.33(18) &1.12(5)& --  & --  &0.86(32)&0.73(3)&0.76(4)\\
2.3158 &                   &                     &--        & --    &0.63(8)& --      & --   & -- & --     & --& --\\
\hline
  \end{tabular}
  \end{minipage}
\end{table*} 

%% file: line_ids_late_rv.tex
\setcounter{table}{1}
\begin{table*}
 \centering
\begin{minipage}{180mm}
  \caption{Line fluxes measured from Mt Abu data. Later data, days 27.76 to 102.66. Fluxes are given in units of $10^{-14}$~W~m$^{-2}$
  and are uncorrected for extinction.
  Uncertainties in the last one, two or three digits are given in brackets; thus 13.0(15) means $13.0\pm1.5$,
  469(11) means $469\pm11$ etc. \label{line_ids_l}}
   \begin{tabular}{cccccccccccc}
  \hline
  $\lambda_{\rm obs}$ & ID, $\lambda_0$ & Transition       & \multicolumn{9}{c}{Line flux ($10^{-14}$~W~m$^{-2}$) on days given}    \\ \cline{4-12}
       ($\mu$m)       &     ($\mu$m)    &   $\ell-u$       & & & & & & & & &  \\ 
   &&&$t=27.76$&31.82&33.75&36.82&37.69&77.79&78.71&101.72&102.66\\ \hline
1.0828&\pion{He}{i}, 1.0833&$^3$S$-^3$P$^o$      & -- & -- &-- & 192(3)&150(2)& -- & 8.57(74)&6.89(16)&6.89(23)\\
1.0936&\pion{H}{i}, 1.0941 & 3--6                & -- & 183(8) & --    &149(3)&135(2)  & -- &-- &1.21(17)&0.35(11)\\
1.0946&\pion{N}{i}, 1.0946 & $^4$S$^o_{3/2}-^4$P & -- & --    & --     & --   & --     & -- & -- & -- &0.36(10)\\
1.1284&\pion{C}{i}, 1.1281 & $^1$D$_2-^1$F$^o_3$ &425(4)&798(9) & -- & -- & 625(5) &1.75 (7)&1.61(7)&0.82(3)&0.71(3)\\
      &\pion{O}{i}, 1.1290 &$^3$P$-^3$D$^o$      & --   & --    & -- & -- & --    &-- & -- & -- & --\\
1.1300&\pion{N}{i}, 1.1297 &$^4$D$^o_{1/2}-^4$P$_{1/2}$
                                                 &--& -- & -- & -- & --& --& --& --& --\\
     &\pion{O}{i}, 1.1301&$^5$P$_2-^5$S$^o_2$    & --    & --  & -- & --& --& --& --& --& --\\
     &\pion{C}{i}, 1.1301&$^3$P$^o_1-^3$D$_2$ & --    & --  & -- & --& --& --& --& --& --\\
      &\pion{N}{i}, 1.1301&$^4$P$^o_{1/2}-^4$P$_{1/2}$  & --    & --  & -- & --& --& --& --& --& --\\
     &\pion{N}{i}, 1.1300&$^4$P$^o_{5/2}-^4$D           & --    & --  & -- & --& --& --& --& --& --\\
     &\pion{N}{i}, 1.1303&$^2$P$^o_{1/2}-^2$D          & --    & --  & -- & --& --& --& --& --& --\\
     &\pion{N}{i}, 1.1304&$^4$D$^o_{3/2}-^4$D$_{1/2}$  & --    & --  & -- & --& --& --& --& --& --\\
1.1463&                   &                            & --    & --  &  -- & --&--& 0.168(55)& --& --& --\\
1.1619&\pion{C}{i}, 1.1618&$^3$D$-^3$D$^o$             & --    &  --  & --  & --& -- & --& -- &-- &--\\
       &\pion{C}{i}, 1.1622&$^3$D$-^3$D$^o$            & --   & --  & --  & --   & --  & --  & --  & --  & --\\
1.1654&\pion{C}{i}, 1.1651&$^3$D$_2-^3$D$^o_1$         &18.8(35)&25.0(18)&23.4(10)& --&17.0 (6) &  -- & --& -- & --\\
1.1662&\pion{C}{i}, 1.1662&$^3$S$-^3$P$^o$             & --&    --& --& --& --& --& --& --& --\\
1.1758&\pion{C}{i}, 1.1758&$^3$D$_2-^3$F$^o_3$        &56.5(31)&73.0(87)&67.5(24)& -- &50.3(53)&0.18(6)&0.31(8)&0.14(3)&--\\
      &\pion{C}{i}, 1.1757&$^3$D$_3-^3$F$^o_4$          & --& --& --& --& --& --& --& --& --\\
1.1863&\pion{C}{i}, 1.1864&$^3$P$_2-^3$D$^o_3$          & -- & --& --& --& --& --& --& --& --\\
      &\pion{C}{i}, 1.1864&$^3$P$_2-^3$D$^o_2$           & -- & -- & --& -- & --& --& --& --& --\\
      &\pion{C}{i}, 1.1866&D$_1-^3$P$^o_1$                & --& --& --& --& --& --& --& --& --\\
      &\pion{N}{i}, 1.1861&$^4$S$^o_{3/2}-^4$P$_{3/2}$ & -- & -- & -- & --& --& --& --& --& --\\
1.1887&                    &                           &11.3(35)&30.7(47)&19.6(11)& -- &11.6(6)& --& --& 0.096(30)& --\\
1.2087&                    &                           & --& -- &18.1(16)& -- &4.34(62)& --& --& --& --  \\
1.2191&                    &                            & --    & --  & -- & --& --& --& --& --& --\\
1.2300  &                   &                            & -- &3.52(32) &--&--&--&--&--&--&--\\
1.2466&\pion{N}{i},1.2473&$^2$D$^o_{5/2}-^2$F$_{7/2}$   &7.24(93)&10.0(12)&14.3(11)&7.35(51)&8.56(18)& --& --& --& \\
1.2509&                  &                                & -- & -- & -- &-- &--&--&--&--&0.20(2)\\
1.2586&\pion{N}{i}, 1.2586&$^2$F$_{7/2}-$D                &11.3(12)&13.0(3)&18.2(14)&9.82(69)&11.0(2)&--&--&--&--\\
      &\pion{C}{i}, 1.2585&$^3$P$_1-^3$P$^o_2$              & -- & --& -- & --& --& --& --& --& --\\
      &\pion{N}{i}, 1.2585&$^2$D$^o_{5/2}-^2$F${5/2}$    & --& --& --& --& --& --& --& --& --\\
1.2809&                   &                              &--&--&--&--&--&1.63(5)&1.77(5)&1.54(4)&1.57(3)\\
1.2823&\pion{H}{i}, 1.2822& 3--5                         &192(2)&230(1)&252(3)&214(2)&251(1)& -- & -- & --& --\\
1.2975&\pion{N}{i}, 1.2973&$^2$P$-^2$D$^o$               &-- & -- & -- & -- & -- & --& --& --& --\\
1.3172&\pion{N}{i}, 1.3177&$^2$D$_{3/2}-^2$D$^o_{5/2}$  &11.1 (11)&8.9(12)&-- &8.42(65)&10.6(2)& --& --& --& --\\
       &\pion{N}{i}, 1.3176&$^2$D$_{5/2}-^2$D$^o_{5/2}$ & -- & --& --& --& --& --& -- & --& --\\
&&&&&&&&&&&\\
1.5268   &--&--                &--      & --    &--     &--&0.42(4)    &--&--&--&--\\
1.5348   &  &                  &1.74(10)&1.4 (1)&1.23(8)&--&0.88(5)&--&--&--&--\\
1.5447   &  &                  &2.45(10)&1.74(10)&1.79(8)& -- &1.29(5)&--&--&--&--\\
1.5568&\pion{H}{i}, 1.5561&4--16&5.68(12)&4.79 (12)&4.28(9)&3.40(7)&3.13(6)&--&--&--&--\\
1.5707&\pion{H}{i}, 1.5705&4--15&4.67(12)&4.03(12)&3.68(9)&2.86(2)&2.18(5)&--&--&--&--\\
1.5748&                   &     &--&--&--&--&--&--&--&--&0.025(4)\\
1.5789&                   &     & --& --& --&-- &--&--&--&--&--\\
1.5847&\pion{H}{i}, 1.5885&4--14&5.44(11)&4.37(11)&4.14(9)& 3.29(7) &3.31(8)&0.17(1)&--&--&0.119(14)\\
1.6021&                   &       &2.53(10)&1.48(10)&1.26(8)&0.90(6)&0.99(7)&--&--&--&--\\
1.6054&                   &         &--&--&--&--&--&--&--&--& 0.027(3)\\
1.6115&\pion{H}{i}, 1.6114&4--13  &6.28(11)&5.30(12)&4.53(9)&3.94(7)&3.52(8)&--&--&--&--\\
1.6142&                   &         &--&--&--&--&--&--&--&--&0.092(16)\\
1.6329&                   &        &--&--&--&--&--&--&--&--&--\\
1.6416&\pion{H}{i}, 1.6412&4--12  &8.60(15)&7.65 (12)&6.93(16)&5.86(56)&5.62(8)&--&--&--&0.183(16)\\
1.6499&                   &        &--&--&0.318(66)&--&--&--&--&--&--\\
1.6711&                   &        &--&--&0.428(72)&--&--&--&--&--&--\\
\hline
  \end{tabular}
  \end{minipage}
\end{table*}

\setcounter{table}{1}
\begin{table*}
 \centering
\begin{minipage}{180mm}
  \caption{Continued. Line fluxes measured from Mt Abu data. Later data, days 27.76 to 102.66. Fluxes are given in units of $10^{-14}$~W~m$^{-2}$
  and  are uncorrected for extinction.
  Uncertainties in the last one, two or three digits are given in brackets; thus 13.0(15) means $13.0\pm1.5$,
  469(11) means $469\pm11$ etc.}
   \begin{tabular}{cccccccccccc}
  \hline
  $\lambda_{\rm obs}$ & ID, $\lambda_0$ & Transition       & \multicolumn{9}{c}{Line flux ($10^{-14}$~W~m$^{-2}$) on days given}    \\ \cline{4-12}
       ($\mu$m)       &     ($\mu$m)    &   $\ell-u$       & & & & & & & & &  \\ 
   &&&$t=27.76$&31.82&33.75&36.82&37.69&77.79&78.71&101.72&102.66\\ \hline
1.6808&\pion{H}{i}, 1.6811&4--11   &10.73(39)&8.90(31)&8.04(26)&7.61(98)&6.81(42)&--&--&--&0.220(24)\\
1.6826&                   &        &--&--&--&--&--&0.327(10)&--&--&--\\
1.6892&\pion{C}{i}, 1.6895&$^1$D$_2-^1$F$^o_3$ &16.9(4)&12.6(3)&11.7(3)&9.59(100)&9.06(42)&--&--&--&0.087(5)\\
1.7058&                   &           &--&--&--&--&--&--&--&--&--\\
1.7247&                   &           &2.02(4)&1.95(14)&1.35(7)&--&--&--&--&--&--\\
1.7362&\pion{H}{i}, 1.7367&4--10       &18.3(6)&17.3(3)&14.1(2)&13.2(7)&13.4(4)&0.276(11)&--&--&0.355(20)\\
1.7450&                  &             &3.46(53)&--&--&--&--&--&--&--&--\\
1.7563&                  &           &--&--&--&--&--&--&--&--&--\\
1.7667&                 &            &--&--&--&--&--&--&--&--&--\\
&&&&&&&&&&&\\
1.9741&                  &           &1.76(9)&1.39(6)&1.24(15)&--&0.863(68)&--&--&--&--\\
2.0599&\pion{He}{i}, 2.0587&$^1$S$_0-^1$P$^o_1$ &1.57 (7)&2.04(16)&4.21(22)&6.67(18)&5.48(7)&--&--&--&0.107(10)\\
2.0757&                     &                   &--&--&--&--&--&--&--&--&--\\
2.0890 &                  &                     &0.51(5)&0.57(4)&0.65(5)&0.47(15)&0.47(7)&--&--&--&--\\
2.1029&\pion{C}{i}, 2.10289&$^1$S$-^1$P$^o$  &0.29(4)&0.221(25)&--&--&--&--&--&--&--\\
2.1235&                   &                  &1.88(7)&1.50 (7)&1.79(10)&1.04(5)&0.718(41)&--&--&--&--\\
2.1492&                  &                   &--&--&--&--&--&--&--&--&--\\
2.1505&                  &                   & --&0.841(7)&--&0.525(53)&0.447(41)&--&--&--&--\\
2.1645&                  &                   &--&--&--&--&--&0.491(24)&--&--&0.366(15)\\
2.1667&\pion{H}{i}, 2.1661&4--7              &15.8(3)&15.7(2)&14.8(2)&14.9(2)&13.0 (1)&--&--&--&--\\
2.2153&                    &                   &--&--&--&--&--&--&--&--&--\\
2.2891&                    &                   &--&--&--&--&--&--&--&--&--\\
2.2918&\pion{C}{i}, 2.2913&$^1$S$_0-^1$P$^o_1$ &--&0.479(22)&--&0.345(31)&0.358(23)&--&--&--&--\\
2.3158 &                   &                     &--        & --    & --& --      & --   & -- & --     & --& --\\
\hline
  \end{tabular}
  \end{minipage}
\end{table*}

%% file: v339del_A1_optical_line_fluxes1.tex
\setcounter{table}{2}
\begin{table*}
 \centering
 \caption{\vd\ optical line fluxes measured from the MMT spectra  on day 672.50. 
 Fluxes  are uncorrected for extinction. \label{line_ids_opt}} 
   \begin{tabular}{cccc}   
$\lambda_{\rm obs}$ & ID, $\lambda_{0}$ & Transition & {Line flux} \\
 {(\AA)} & {(\AA, air)} & {$l-u$} & ($10^{-17}$ W m$^{-2}$) \\ \hline
3963.381 & H$\epsilon$, 3970.072 & $2 - 7$ & 10.04\\
3972.851 & H$\epsilon$, 3970.072 & $2 - 7$ & 5.90 \\ 
4094.305 & H$\delta$, 4101.135 & $2 - 6$ & 21.77\\
4105.861 & H$\delta$, 4101.135 & $2 - 6$ & 14.17 \\
4261.049 & \pion{C}{ii}, 4267.183 & $^2$D$-^2$F$_{0}$ & 3.38 \\
4273.072 & \pion{C}{ii}, 4267.183 & $^2$D$-^2$F$_{0}$ & 2.26 \\
4334.038 & H$\gamma$, 4340.463 & $2 - 5$ & 23.40 \\
4354.927 & H$\gamma$, 4340.463 & $2 - 5$ & 29.52 \\
4344.003 & \fion{O}{iii}, 4363.209 & $^1$D$-^1$S & 11.02 \\
4370.276 & \fion{O}{iii}, 4363.209 & $^1$D$-^1$Ss & 16.74 \\
4632.792 & \pion{N}{iii}, 4640.640 & $^2$P$_{0}-^2$D & 18.30 \\
4646.433 & \pion{N}{iii}, 4640.640 & $^2$P$_{0}-^2$D & 13.77 \\
4678.127 & \pion{He}{ii}, 4685.710 & $3 - 4$ & 12.07 \\
4691.931 & \pion{He}{ii}, 4685.710 & $3 - 4$ & 8.66 \\
4853.992 & H$\beta$, 4861.325 & $2 - 4$ & 54.77 \\
4867.393 & H$\beta$, 4861.325 & $2 - 4$ & 38.87 \\
4951.556 & \fion{O}{iii}, 4958.911 & $^3$P$-^1$D & 482.3 \\
4965.698 & \fion{O}{iii}, 4958.911 & $^3$P$-^1$D & 320.8 \\
4999.424 & \fion{O}{iii}, 5006.843 & $^3$P$-^1$D & 1453.0 \\
5013.692 & \fion{O}{iii}, 5006.843 & $^3$P$-^1$D & 973.0 \\
5168.903 & \fion{Fe}{vi}, 5176.040 & $^4$F$-^2$G & 2.25\\
5191.028 & \fion{Fe}{vi}, 5176.040 & $^4$F$-^2$G & 2.46\\
5405.475 & \pion{He}{ii}, 5411.520 & $4 - 7$ & 1.18\\
5418.364 & \pion{He}{ii}, 5411.520 & $4 - 7$ & 1.08\\
5526.680 & \pion{Ar}{ii}, 5534.990 & $^4$D$-^4$P$_{0}$ & 0.92\\
5541.127 & \pion{Ar}{ii}, 5534.990 & $^4$D$-^4$P$_{0}$ & 0.69\\
5670.579 & \fion{Fe}{vi}, 5676.950   & $^4$F$-^4$P & 4.21 \\
5685.666 & \fion{Fe}{vi}, 5676.950   & $^4$F$-^4$P & 2.47 \\
5745.688 & \fion{N}{ii}, 5754.644    & $^1$D$-^1$S & 6.66 \\
5762.683 & \fion{N}{ii}, 5754.644    & $^1$D$-^1$S & 4.14 \\
5866.875 & \pion{He}{i}, 5875.966    & $^3$P$_{0}-^3$D & 6.93 \\
5882.082 & \pion{He}{i}, 5875.966    & $^3$P$_{0}-^3$D & 5.31 \\
5932.956 & \pion{N}{i}, 5931.780     & $^3$P$-^3$D$_{0}$ & 0.91 \\
5948.923 & \pion{N}{i}, 5931.780     & $^3$P$-^3$D$_{0}$ & 0.37 \\
6076.424 & \fion{Ca}{v}, 6086.400 + \fion{Fe}{vii}, 6086.290 & $^3$P$-^1$D; $^3$F$-^1$D & 1.20 \\
6096.129 & \fion{Ca}{v}, 6086.400 + \fion{Fe}{vii}, 6086.290 & $^3$P$-^1$D; $^3$F$-^1$D & 1.09 \\
6292.857 & \fion{O}{i}, 6300.304     & $^3$P$-^3$D & 26.10 \\
6307.521 & \fion{O}{i}, 6300.304     & $^3$P$-^3$D & 16.46 \\
6356.264 & \fion{O}{i}, 6363.766 + \fion{Fe}{x}, 6374.500 & $^3$P$-^1$D; $^2$P$_{0}-^2$P$_{0}$ & 8.93 \\
6371.086 & \fion{O}{i}, 6363.766 + \fion{Fe}{x}, 6374.500 & $^3$P$-^1$D; $^2$P$_{0}-^2$P$_{0}$ & 5.55 \\
6536.497 & \fion{N}{ii}, 6548.040 & $^3$P$-^1$D & 111.60\\
6559.453 & \fion{N}{ii}, 6548.040 & $^3$P$-^1$D & 153.50\\
6549.564 & H$\alpha$, 6562.80 + \pion{He}{ii}, 6560.10 &$2 - 3; 4 - 6$ & 221.30\\
6582.781 & H$\alpha$, 6562.80 + \pion{He}{ii}, 6560.10 &$2 - 3; 4 - 6$ & 305.00\\
6571.919 & \fion{N}{ii}, 6583.460 & $^3$P$-^1$D & 458.20\\
6594.550 & \fion{N}{ii}, 6583.460 & $^3$P$-^1$D & 239.90\\
6669.174 & \pion{He}{i}, 6678.152    & $^1$P$_{0}-^1$D & 2.23 \\
6685.912 & \pion{He}{i}, 6678.152    & $^1$P$_{0}-^1$D & 1.83 \\
7054.257 & \pion{He}{i}, 7065.215    & $^3$P$_{0}-^3$D & 1.92 \\
7071.989 & \pion{He}{i}, 7065.215    & $^3$P$_{0}-^3$D & 1.68 \\ \hline
\end{tabular}
\end{table*}
